\documentclass[preprint,11pt]{elsarticle}

\usepackage[top=2.5cm,bottom=2.5cm,left=2.6cm,right=2.6cm,a4paper]{geometry}
\usepackage{graphicx}
\usepackage{dcolumn}
\usepackage{bm}
\usepackage{epstopdf}
\usepackage{mathrsfs}
\usepackage{amssymb,amsfonts,latexsym}
\usepackage{amsmath,bbold}
\usepackage{color}

\def\bs{\boldsymbol}
\def\del{\partial}
\def\bdel{\bs\partial}

\long\def\comment#1{ }

\def\p{{\boldsymbol p}}

\def\0{{\boldsymbol 0}}
\def\q{{\bm q}}
\def\l{{\boldsymbol l}}
\def\k{{\boldsymbol k}}

\def\x{{\boldsymbol x}}
\def\y{{\boldsymbol y}}

\def\r{{\boldsymbol r}}
\def\z{{\boldsymbol z}}
\def\u{{\boldsymbol u}}
\def\v{{\boldsymbol v}}

\def\SII{S^{(2)}}
\def\SIII{S^{(3)}}

\def\Gc{{\cal G}}
\def\tiSIII{{\tilde S}^{(3)}}
\newcommand{\beq}{\begin{eqnarray}}
\newcommand{\eeq}{\end{eqnarray}}
\newcommand{\be}{\begin{eqnarray*}}
\newcommand{\ee}{\end{eqnarray*}}
\newcommand{\bal}{\begin{align}}
\newcommand{\eal}{\end{align}}
\newcommand{\rmd}{{\rm d}}
\newcommand{\rme}{{\rm e}}

\newcommand{\Romega}{\omega'}
\newcommand{\nn}{\nonumber\\ }

\begin{document}
\begin{frontmatter}

\title{Renormalization of the jet-quenching parameter}
\author[ipht]{Jean-Paul Blaizot}
\author[ipht]{Yacine Mehtar-Tani}

\address[ipht]{%
Institut de Physique Th\'eorique, CEA Saclay, F-91191 Gif-sur-Yvette,
France
}%


\begin{abstract}
We study the radiative processes that affect the propagation of a high energy gluon in a dense medium, such as a quark-gluon plasma. In particular, we investigate the role of the  large double logarithms corrections, $\sim\alpha_s \ln^2 L/\tau_0$,  that were recently identified in the study of $p_\perp$-broadening by Liou, Mueller and Wu.  We show that these large  corrections can be reabsorbed in a renormalization of the jet quenching parameter controlling  both momentum broadening and energy loss. We argue that the probabilistic description of these phenomena remains valid,  in spite of the large non-locality in time of the radiative corrections.  The renormalized jet-quenching parameter is enhanced compared to its standard perturbative estimate. As a particular consequence,  the radiative energy loss scales with medium size $L$ as $L^{2+\gamma}$, with $\gamma=2\sqrt{\alpha_s N_c /\pi}$, as compared to the standard  scaling in $L^2$. 
\end{abstract}


\end{frontmatter}
\begin{flushright}
\end{flushright}

\section{Introduction}

A high energy parton propagating through a dense medium, cold nuclear matter, or a high temperature quark-gluon plasma, undergoes multiple interactions with the constituents of the medium.  Among those are the elastic scatterings, leading to the broadening of the transverse momentum of the parton with respect to the direction of its initial momentum. In addition, the energetic parton may experience medium-induced soft gluon radiations that degrade its energy. These two important processes, momentum broadening and energy loss,  are governed by a transport coefficient, called the jet quenching parameter, and denoted traditionally by $\hat q$ \cite{Baier:1996kr}.  The jet-quenching parameter also controls the splitting of gluons along the in-medium QCD cascade \cite{Blaizot:2013vha}. It has become a central concept in the analysis of jet observables in Heavy-Ion Collisions \cite{d'Enterria:2009am,Mehtar-Tani:2013pia}.

The description of these phenomena in the framework of perturbative QCD is well understood from  the  BDMPS-Z theory  \cite{Baier:1996kr,Baier:1996sk,Zakharov:1996fv,Zakharov:1997uu}. Let us recall that, at leading order, the medium-induced gluon radiation spectrum reads  (for gluon frequencies $\omega\lesssim \omega_c\equiv \hat q L^2$),
\beq\label{BDMPS}
\omega\frac{\rmd N}{\rmd \omega}\simeq \frac{\alpha_s C_R}{\pi} \sqrt{\frac{\hat q }{\omega}}\,L\,,
\eeq 
where $C_R$ is the color charge of the emitter, $C_A=N_c$ for a gluon and $C_F=(N_c^2-1)/2N_c$ for a quark, respectively, and $L$ is the typical distance travelled in the medium by the high energy parton. The jet-quenching parameter measures the amount of transverse momentum squared that is acquired by the parton per unit length traversed in the medium.  That the spectrum (\ref{BDMPS}) depends on the same jet-quenching parameter reflects the basic process at work in medium-induced radiation: the matching between the formation time of a gluon $\tau_f\sim \omega/k_\perp^2$, and the transverse momentum acquired during $\tau_f$, $k_\perp^2\sim \hat q \,\tau_f$, which determines the  typical branching time as $\tau_f\sim \tau_\text{br}\equiv \sqrt{\omega/\hat q}$, and  the spectrum above, $\omega\rmd N/\rmd\omega\propto L/\tau_{\rm br}$. It follows  from this spectrum that the mean energy lost through radiation by a particle going through a medium is given (parametrically) by  $\langle \omega\rangle \simeq \alpha_s C_R \,\hat q\, L^2 $. Note that this average is dominated by hard but rare emissions, with frequencies near the cut-off, $\omega\sim \omega_c$. 

The jet-quenching parameter $\hat q$ introduced above is usually interpreted as a transport coefficient. As such, it depends in a complicated way on the properties of the medium, and may receive genuine non-perturbative contributions \cite{Majumder:2012sh,Panero:2013pla}. However, the entire description of the propagation of a fast parton in a dense medium may also be affected by another type of contributions, namely radiative corrections. Such  corrections to the typical transverse momentum broadening $ \langle k_\perp^2\rangle_\text{typ}$, were only recently considered by Liou, Mueller and Wu \cite{Wu:2011kc,Liou:2013qya} who have identified potentially large double logarithmic contributions $\sim \alpha_s \ln^2 L/\tau_0$, with typically $L$ the length of the medium and $\tau_0\sim 1/T$ (in a quark-gluon plasma with temperature $T$)\footnote{Radiative corrections to $p_\perp$-broadening are being also investigated in the higher twist formalism \cite{Kang:2013raa}.}.   It is an interesting question to study whether these corrections possess some universal character, that is, whether they imply a modification of the jet quenching parameter that controls the momentum broadening, and whether that modification is the same independently of the process where $\hat q$ enters, in particular, whether the same correction to $\hat q$ modifies the radiation spectrum.

This paper builds up on our previous paper \cite{Blaizot:2013vha}, where the double logarithmic correction emerged as a correction to the transport equation describing parton propagation, and hence naturally appeared as a correction to the jet quenching parameter.  However, the approach followed in Ref.~\cite{Blaizot:2013vha} leaves  unanswered the question  alluded to above regarding the universality of the correction to $\hat q$.  To answer this question is the main motivation for this paper. To that aim, we shall explicitly calculate the dominant radiative correction to the emission kernel that controls gluon splitting, and hence other phenomena such as the energy loss, or more generally the in-medium gluon cascade. We shall show  here that the double logarithmic correction  renormalizes in the same way the jet-quenching parameters that control the momentum broadening and
the radiation spectrum.  A similar result hods in more general cases, in particular for the QCD cascade, at least in the limit of large $N_c$.

The outline of this paper is as follows. 
In Sect.~\ref{pt-broad}, we put in place elements of the formalism that we use, by reviewing various aspects of the calculation of the $p_\perp$-broadening. In Section \ref{rad-pt-broad}, we consider the first radiative correction to the broadening probability that is enhanced by a double logarithm. We argue that  the logarithmic nature of the phase-space allows for a simple exponentiation of the single radiative correction, in spite of its non locality in time. This enables us to interpret the   double logarithmic correction as a renormalization of the jet-quenching parameter. Then, in Section \ref{rad-E-loss}, we show that the same correction to $\hat q$ applies to the medium-induced gluon spectrum, confirming the universality of the  renormalization. We show that, in the limit of large $N_c$,  this result extends to the splitting kernel, the building block of the in-medium QCD cascade.  Finally, we discuss the evolution of $\hat q$ that results form a resummation of the leading radiative corrections\footnote{A possible evolution of the jet-quenching parameter was first discussed in \cite{CasalderreySolana:2007sw}.}, and we briefly comment on possible implication for the phenomenology of the jet-quenching. Throughout this paper, we restrict our discussion to gluon jets. The generalization to quark jets is straightforward.

\section{$p_\perp$-broadening in the approximation of independent multiple scatterings}\label{pt-broad}
In this section, we put in place various elements of the formalism that we shall use throughout this paper. We do so by reviewing the calculation of the $p_\perp$-broadening probability distribution in the approximation of independent multiple scatterings \cite{Gyulassy:1993hr,Wiedemann:2000za,MehtarTani:2006xq}. We consider a high energy gluon moving in the $+z$ direction, and  model the medium in which it propagates as a fluctuating random field $A^-(x^+,\x)$ that depends only on the light-cone time $x^+$ and the transverse coordinates  $\x$.\footnote{We use light-cone coordinates $x^{\pm}=(t\pm z)/\sqrt{2}$} We treat the interaction with the medium in the  eikonal approximation. In particular, the polarization of gluons is unchanged during their propagation through the medium. Furthermore, the field being independent of $x^-$, the $+$ component of the momentum is also conserved. Both the $+$ momentum and the polarization will be omitted to alleviate  the notation

The amplitude for a gluon of (large) momentum $p_0\equiv(p^+_0,\p_0=0)$, 
present in the system at time $t_0$, to evolve in the medium into a gluon with momentum $p_1=(p^+_1,\p_1)$, where $p^+_0=p^+_1\equiv E$, 
is given by (to within an irrelevant phase factor)
\beq\label{Mone}
{\cal M}(p_1|p_0)=  
(\p_1|\,{\cal
G}(t_1,t_0)\,|\p_0)\,.
\eeq
 The amplitude ${\cal M}\equiv {\cal M}_{ab}$ is a matrix in color space, propagating a gluon with color $b$ to color $a$. 
We denote the light--cone time $x^+$ simply as $t$, to simplify the notation. (Note that $t_1-t_0=\sqrt{2} L$, where $L$ is the longitudinal length of the medium.)

The propagator ${\cal G}$ is that of a Schr\" odinger equation
in  two dimensions (the transverse plane) for a non-relativistic particle of mass $E$ 
moving in a time-dependent potential $A^-(t,\x)$.
That is, it satisfies the equation
\beq\label{Schroedinger}
\left[ i\del_t +g(T\cdot A^-)+\frac{\nabla_\perp^2}{2E}\right]_{ac}(\x|\,{\cal G}^{cb}(t,t_0)\,|\x_0)=i\delta_{ab}\delta(t-t_0)\delta(\x-\x_0)\,.
\eeq
where $T\cdot A^-\equiv T^a\cdot A_a^-$ and $T^a$ are the generators of $SU(3)$ in the adjoint representation. 
The solution can be written as  a path integral
\beq\label{path-int}
(\x_1|\,{\cal G}(t_1,t_0)\,|\x_0)=\theta(t_1-t_0)\int {\cal D} \r\, \exp\left(i\frac{E}{2} \int_{t_0}^{t_1}    \rmd t \, \dot\r^2  \right)\, U_\r (t_1,t_0)\,,
\eeq
with $\r(t_1)=\x_1$, $\r(t_0)=\x_0$, $\dot \r=\del_t \r$,  and $U_\r$ is a Wilson-line  in the adjoint representation evaluated along the path $\r(t)$:
\beq
U_\r (t_1,t_0)={\rm T}\exp\left[ ig\int_{t_0}^{t_1} \rmd t \, A^-(t, \r(t))\cdot\, T   \right]\,.
\eeq

An alternative way of writing Eq.~(\ref{Schroedinger}) is as an integral equation:
\beq\label{integralequation1}
{\cal G}(t_1,t_0)={\cal G}_0(t_1,t_0)+\int_{t_0}^{t_1} {\rm d} t\, {\cal G}_0(t_1,t) \, [igA^-(t)\cdot T]\, {\cal G}(t,t_0)\,,
\eeq
where we use a matrix notation for ${\cal G}$ (in both color and coordinate spaces). The free propagator ${\cal G}_0$ can be written as  
\beq\label{freecalGx}
(\x_1|{\cal G}_0(t_1-t_0)|\x_0)=\theta(t_1-t_0)\left(\frac{E}{2\pi i\Delta t}   \right)\, \exp\left[i\frac{E}{2\Delta t}(\x_1-\x_0)^2\right]\,,
\eeq
with $\Delta t\equiv t_1-t_0$. Note that the typical value of $\Delta x_\perp=|\x_1-\x_0|$ reached after a time $\Delta t$ is $\Delta x_\perp\sim \sqrt{2\Delta t/E}$. Note also that, as  $\Delta t/E\to 0$, $(\x_1|{\cal G}_0(\Delta t)|\x_0)\to \delta(\x_1-\x_0)$, and the full propagator (\ref{path-int}) reduces to 
\beq\label{path-int2}
(\x_1|{\cal G}(t_1,t_0)|\x_0)=\theta(t_1-t_0) \delta(\x_1-\x_0) \, U_{\x_0}(t_1,t_0),
\eeq
where $U_{\x_0}(t_1,t_0)$ is a trivial Wilson line along the straight path $\x_0={\rm cste}$.

In general,  the propagator ${\cal G}$ is easier to evaluate in coordinate space, where it is given explicitly, for instance, by the path integral (\ref{path-int}). However the physical amplitude that we are interested in has a natural expression in momentum space. Indeed, the quantity that we wish to calculate is the probability that the initial gluon acquires a given transverse momentum through its interaction with the medium. This is obtained by squaring the amplitude,  averaging over the initial color and polarization, summing over final polarization and color, and finally averaging over the background field. In doing this calculation, it is convenient to keep the initial momentum $\p_0$ in the amplitude distinct from that, $\bar\p_0$, in the complex conjugate amplitude. We get then
\beq\label{defS2kk}
{\cal M}(\p_1|\p_0)\,{\cal M}^*(\p_1|\bar\p_0) &=&\frac{1}{N_c^2-1}\left<{\rm Tr}
(\p_1|\,{\cal G}(t_1,t_0)\,|\p_0)(\bar\p_0|\,{\cal G}^\dagger(t_0,t_1)\,|\p_1)\right>\nn
&\equiv& (\p_1;\p_1|\,S^{(2)}(t_1,t_0)\,|\p_0;\bar\p_0)\,,\nn
&=&(2\pi)^2 \delta(\bar\p_0-\p_0) \, {\cal P}(\p_1,t_1|\p_0,t_0)\,,
\eeq
where the trace is over color indices and in the last line we have factored out the momentum conserving delta function that reflects translational invariance. The angular brackets denote medium average, to be specified shortly. The quantity ${\cal P}(\p_1,t_1|\p_0,t_0)$  can be interpreted as the $p_\perp$-broadening probability distribution. It is easily checked in particular that $\int_{\p_1}{\cal P}(\p_1,t_1|\p_0,t_0)=1$. Here we introduce a shorthand notation for transverse momentum integrations, to be used thoughout the paper: $\int_\q\equiv \int {\rm d}^2\q/(2\pi)^2$.

\begin{figure}[htbp]\label{2-point}
\begin{center}
\includegraphics[width=7cm]{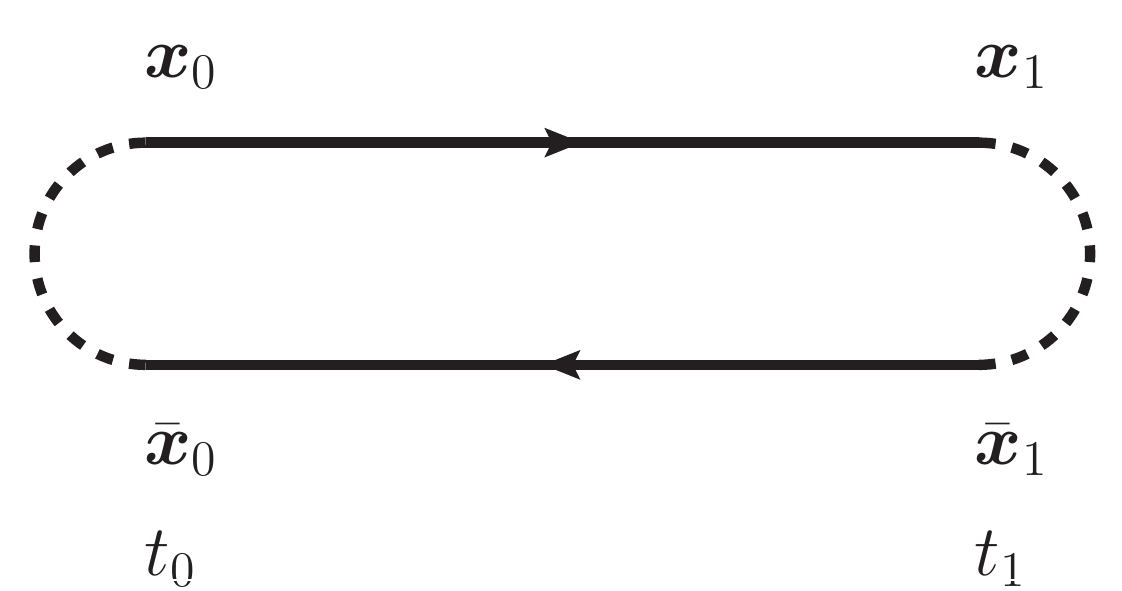}
\caption{Diagrammatic representation of the 2-point function. }
\end{center}
\end{figure}

\begin{figure}[htbp]
\begin{center}
\includegraphics[width=7cm]{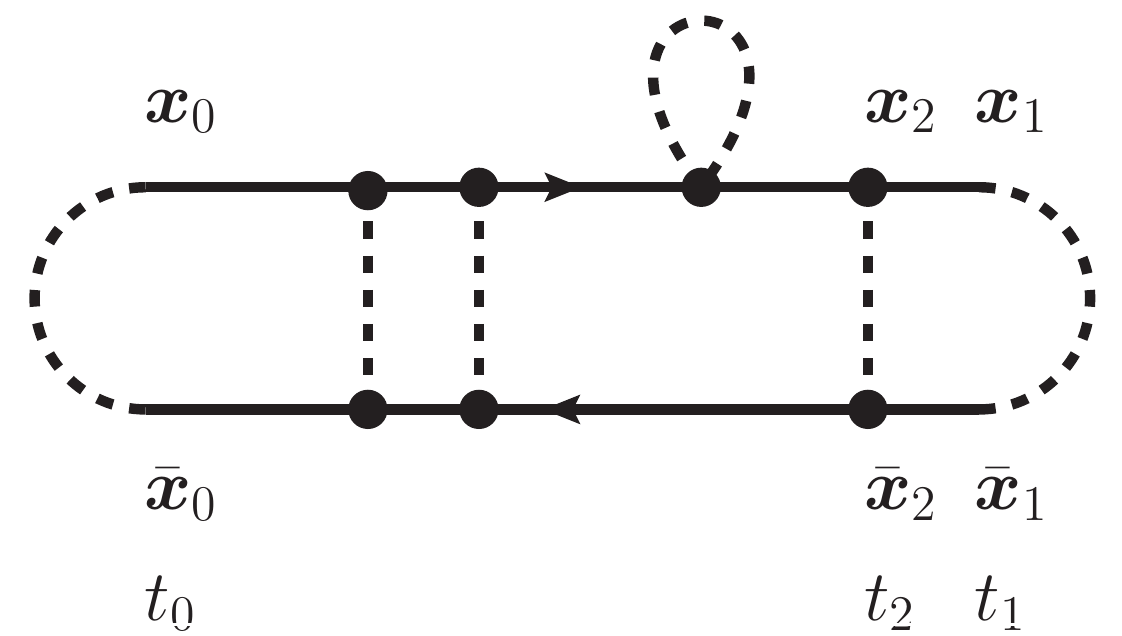}
\caption{A contribution to the 2-point function after averaging over the medium configurations. The medium averaging accounts for collisions with the medium constituents, and involves  instantaneous interactions (vertical dotted lines) or local self-energy insertions (the dotted loop). These two types of contributions will be often referred to later, in a broader context, as real and virtual contributions, respectively.}
\label{fig:fig2}
\end{center}
\end{figure}

The quantity $S^{(2)}$ introduced in Eq.~(\ref{defS2kk}) will be referred to as a 2-point function \cite{Blaizot:2012fh}. Its expression in coordinate space will be useful. This is given, quite generally (see Fig.~\ref{fig:fig2} for an illustration), by  
\beq\label{2-pt-def}
(X_1|\SII(t_1,t_0)|X_0)=\frac{1}{N_c^2-1}\langle \text{Tr}~  (\x_1|\Gc(t_1,t_0)|\x_0)\,(\bar\x_0|\Gc^\dag(t_0,t_1)|\bar\x_1)\rangle\,.
\eeq
where $X_i=(\x_i;\bar\x_i).$\footnote{Some explanation about the notation is needed here. We regard $S^{(2)}(t_1,t_0)$ as a matrix in transverse space, with matrix elements in coordinate space given by $(Y|S^{(2)}(t_1,t_0)|X)$, where $X=(\x;\bar\x)$, $Y=(\y;\bar\y)$ and  the coordinates $\x,\bar\x,\y,\bar\y$ attached to ${\cal G}$ and ${\cal G}^\dagger $ as indicated in Eq.~(\ref{2-pt-def})  (the time variables are not considered as matrix indices, and $S^{(2)}(t_1,t_0)$ is proportional to $\theta(t_1-t_0$)). The same notation applies to the momentum space representation, and we could for instance write the second line  in Eq.~(\ref{defS2kk}) as $(P_1|\SII(t_1,t_0)|P_0)$, with $P_0=(\p_0;\bar \p_0)$, $P_1=(\p_1;\p_1)$, and the semi-colon separating the variables in the amplitude and the complex conjugate amplitude.}
As mentioned earlier the medium is model by a random field $A^-$. The average over the field configurations is assumed to be gaussian with a correlator given by  
\beq\label{AA}
\left< A^-_a(\x,t)A^-_b(\y,t')    \right>=\delta_{ab} \,n\,\delta(t-t')\gamma(\x-\y)\,,
\eeq
where $n$ is the density of color charges (which we assume to be independent of $x^+=t$ for simplicity) and  
\beq\label{gamma}
\gamma(\x)=g^2\int_\q \frac{e^{i\q\cdot\x}}{\q^4}\,.
\eeq

With the 2-point correlator given by Eq.~(\ref{AA}), and the path integral representation (\ref{path-int}) of the propagator, we can easily perform the gaussian average over the random field $A^-$. One may then obtain an explicit expression for $S^{(2)}$ in terms of a path integral. Alternatively, $S^{(2)}$ may be written as the solution of the following integral equation, that is easily obtained from Eq.~(\ref{integralequation1}):
\beq\label{integralequation2}
S^{(2)}(t_1,t_0)=S^{(2)}_0(t_1,t_0)-\int_{t_0}^{t_1} {\rm d}t_3\,\int_{t_0}^{t_3} {\rm d}t_2\, S^{(2)}_0(t_1,t_3)\,\Sigma^{(2)}(t_3,t_2) \, S^{(2)}(t_2,t_0)\,.
\eeq
Here the quantity $\Sigma^{(2)}$ is the matrix, diagonal in coordinate space, defined by \footnote{The delta function $\delta(t-t') $ reflects the instantaneity of the correlator (\ref{AA}).}
\beq\label{sigmaSigma2}
(X|\Sigma^{(2)}(t,t')|Y)\equiv \delta(X-Y) \delta(t-t')\frac{N_c}{2}n\,\sigma(\x-\bar\x)\,,
\eeq
where $\delta(X-Y)=\delta(\x-\y)\delta(\bar\x-\bar\y)$, and $\sigma(\x)$ is the so-called dipole-cross section 
\beq\label{dipole-cross}
\sigma(\x)=2g^2\left[\gamma(\0)-\gamma(\x)\right]\,.
\eeq

In the  high energy limit,  the changes in the coordinates $\x,\bar\x$ of the dipole from $t_0$ to $t_1$ can be neglected (eikonal approximation). Taking  the limit $\Delta t/E\to 0$ in Eq.~(\ref{integralequation2}), and recalling Eq.~(\ref{path-int2}), one gets (with $\v=\x_0-\bar\x_0$, $\delta(X_1-X_0)=\delta(\x_1-\x_0)\delta(\bar\x_1-\bar\x_0)$)
\beq\label{integralequation3}
(X_1|S_\text{eik}^{(2)}(t_1,t_0)|X_0)=\delta(X_1-X_0)-\frac{N_c}{2}n\int_{t_0}^{t_1} {\rm d}t \,\sigma(\v) \, (X_1|S_\text{eik}^{(2)}(t,t_0)|X_0),
\eeq
whose solution reads
\beq\label{S2eikonal}
(X_1|S_\text{eik}^{(2)}(t_1,t_0)|X_0)= \delta(X_1-X_0)\exp\left[-\frac{N_c}{2}n \,(t_1-t_0)
\, \sigma(\v)\right]\,.
\eeq
The exponential factor is commonly interpreted as the $S$-matrix for the elastic scattering of a color dipole of (fixed) transverse size $\v$ \cite{Liou:2013qya}. The instantaneous interactions of the dipole with the medium exponentiate, leading to the interpretation of the dipole $S$-matrix in terms of multiple scatterings. This exponentiation may also be viewed as a diagram resummation.  An illustration of the color averaging and the corresponding diagrams that are resummed is given in  Fig.~\ref{fig:fig2}. Anticipating on the forthcoming discussion, let us observe here that the exponentiation results from the instantaneous nature of the  interactions and the causal  propagation between successive interactions.

By taking the Fourier transform of Eq.~(\ref{integralequation2}), one gets
an equation for ${\cal P}$:
\beq\label{ME}
{\cal P}(\p_1,t_1|\p_0,t_0)=(2\pi)^2\delta(\p_1-\p_0)-\frac{N_c}{2}n\int_{t_0}^{t_1} {\rm d}t \int_{\q} \sigma(\q)\,{\cal P}(\p_1-\q,t|\p_0,t_0)\,,\nn
\eeq
where $
\sigma(\q)=\int d^2\v\, {\rm e}^{i\q\cdot\v}\sigma(\v)$ is the Fourier transform of the dipole cross-section. 
This equation has as simple probabilistic interpretation. To see that more clearly, it is useful to write, using Eqs.~(\ref{dipole-cross}) and (\ref{gamma}),
\beq\label{sigma-w}
-\frac{N_c}{2}n\sigma(\q)\equiv w(\q)-(2\pi)^2\delta(\q)\int_{\q'} w(\q')\,,\qquad w(\q)=\frac{nN_cg^2}{\q^4}\,.
\eeq
The equation (\ref{ME}) can then be written as a master equation
\beq
{\cal P}(\p_1,t_1|\p_0,t_0)&=&(2\pi)^2\delta(\p_1-\p_0)\nn &+& \int_{t_0}^{t_1} {\rm d}t \int_{\q} w(\q){\cal P}(\p_1-\q,t|\p_0,t_0)-\int_{t_0}^{t_1} {\rm d}t \,{\cal P}(\p_1,t|\p_0,t_0)\int_{\q} w(\q)\,,\nn
\eeq
in which 
$w(\q) \rmd t$ can be interpreted as the probability that the particle acquire a transverse momentum $\q$ during $\rmd t$, and the two terms in the last line can be viewed as respectively a gain and a loss term. Note that $w(\q)$ is proportional to the elastic scattering cross section $\rmd^2\sigma_\text{el}/\rmd^2\q$ of the fast parton with the constituents of the medium.
The integral equation (\ref{ME}) can be rewritten as a differential equation by taking the derivative with respect to $t_1$ on both sides. One gets, with a slight change of notations (${\cal P}(\p,t|\p_0,t_0)\equiv{\cal P}(\p,t)$),
\beq
\frac{\del}{\del t}{\cal P}(\p,t)=-\frac{N_c }{2}n\int_\q\sigma(\q){\cal P}(\p-\q,t)\,. 
\eeq
In the regime dominated by multiple soft scatterings, namely when the momentum transfer in a collision with medium particles is small compared to the typical momentum acquired at time $t$, i.e., when $\q\ll \p$, one can transform this equation into a Fokker-Planck equation: 
\beq\label{P-mom-FP}
\frac{\del}{\del t}{\cal P}(\p,t) = \frac{1}{4}\,\left(\frac{\del}{\del \p}\right)^2\,\left[ \hat q(\p^2) \,{\cal P}(\p,t)\right]\,,
\eeq
with the (momentum dependent) diffusion coefficient given, to logarithmic accuracy,  by 
\beq\label{qhat0}
\hat q(\p^2) = \int_\q \q^2 w(\q)=-\frac{N_c}{2} n\int_\q \q^2 \sigma(\q) \simeq 4\pi \alpha_s^2 N_c \,n  \, \ln\frac{\p^2}{m_D^2}\,.
\eeq
The $\p^2$ dependence of $\hat q$ comes from the integration of the logarithmic phase-space of $\q^2$ from $m_D^2$ to $\p^2$, with $m_D $ the Debye screening mass. Note that only the part $w(\q)$ of $\sigma(\q)$ (see Eq.~(\ref{sigma-w})), that is, only  the ``real'' term,  contributes to $\hat q$ (the virtual term does not involve any transfer of momentum).
Note also that the transport coefficient determines the expansion of the dipole-cross section for small sizes, the so-called harmonic approximation, 
\be\label{qhat-r2}
\frac{N_c}{2} n \sigma(\v)\simeq \frac{1}{4}\hat q(\v^{-2}) \, \v^2\,.
\ee
This is obtained by expanding the dipole cross section (\ref{dipole-cross}) for small $\v$, and using $1/\v$ as upper cutoff in the logarithmic integral giving $\hat q$. This expression will be referred to later in the discussion.

\section{Radiative corrections to the broadening probability}\label{rad-pt-broad}
Let us turn now to radiative corrections to the broadening probability or equivalently to the 2-point function. 
The results to be presented in this section are not new (see e.g. \cite{Liou:2013qya,Blaizot:2013vha}). However, it is instructive, and useful for the following section, to review the basic steps in their derivation, in order to exhibit some of the subtleties of the calculation. We shall first discuss the correction to the broadening probability distribution, and then briefly rephrase the calculation in coordinate space. Both versions contain useful and complementary information.  
Additional technical details are given in the Appendix. 

\begin{figure}[htbp]
\begin{center}\label{3-point}
\includegraphics[width=6cm]{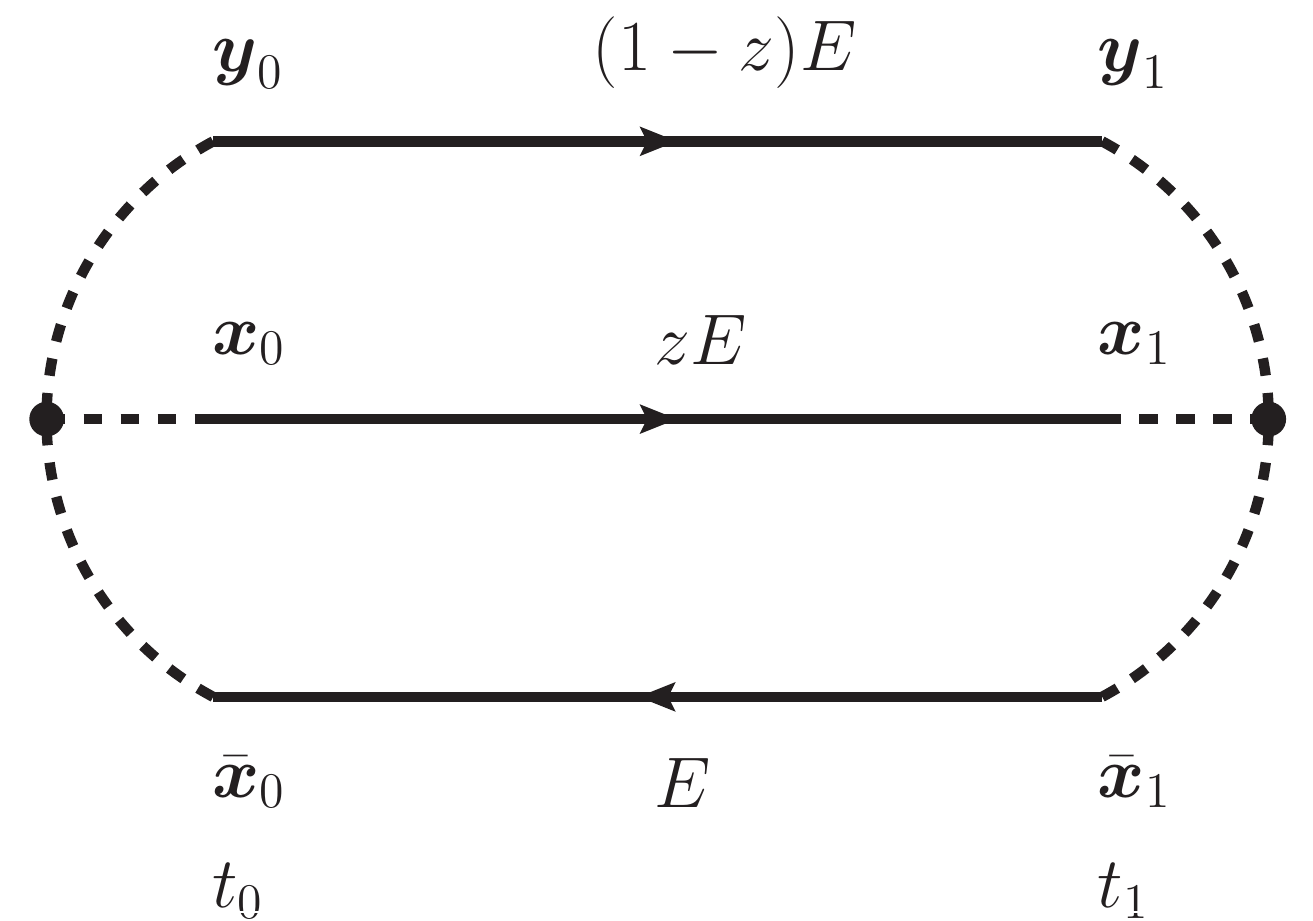}
\caption{Diagrammatic representation of the 3-point function $(X_1|S^{(3)}(t_1,t_0)|X_0)$ associated to the emission of a soft gluon \cite{Blaizot:2012fh}. Here   $X_0$ and $X_1$ denote collectively the transverse coordinates attached to the end points of the propagators: $X_0=(\x_0,\y_0;\bar \x_0)$, and $X_1=(\x_1,\y_1;\bar \x_1)$. The middle line represents the propagator ${\cal G}$ of the hard gluon after its splitting, in the amplitude, and the bottom line the propagator ${\cal G}^\dagger$ of the hard gluon in the complex conjugate amplitude. The splitting occurs at time $t_0$ in the amplitude, and at time $t_1>t_0$ in the complex conjugate amplitude. The upper line represents the propagator ${\cal G}$ of the soft emitted gluon ($1-z\ll 1$). In the eikonal approximation, the propagators associated with the hard gluon are simply trivial Wilson lines $U^\dagger_{\bar \x_0}(t_1,t_0)$ and $U_{\x_0}(t_1,t_0)$.} 
\end{center}
\end{figure}

Before we proceed with radiative corrections,  we need to introduce the 3-point function that plays a central role in the calculation, and review some of its basic properties. 
This 3-point function is defined in analogy with the 2-point function of Eq.~(\ref{2-pt-def}). It can be written in general as follows, see Fig.~\ref{3-point},
\beq\label{3-pt-def}
(X_1|\SIII (t_1,t_0)|X_0)=\frac{\,f^{a_1b_1c_1}f^{a_0b_0c_0}}{N_c(N_c^2-1)}\,\langle (\y_1|\Gc^{a_1a_0} |\y_0) (\x_1|\Gc^{b_1b_0} |\x_0)(\bar\x_0|\Gc^{\dagger c_0c_1} |\bar\x_1)\rangle\,, 
\eeq
using a similar matrix notation as for the 2-point function, but with here $X_0\equiv(\x_0,\y_0;\bar\x_0)$ and similarly for $X_1$. In the right hand side, we have not indicated the dependence of the propagators on $t_1$ and $t_0$ to simplify the writing. This 3-point function involves the product of three gluon propagators, two ${\cal G}$ and one ${\cal G}^\dagger$, the whole object being coupled to an overall color singlet.  An explicit expression analogous to that given above for the 2-point function in terms of the dipole cross section can be obtained by using the path integral representation of the propagator (see e.g. \cite{Blaizot:2012fh}), and averaging over the background field. One then easily derives the following integral equation, analogous to Eq.~(\ref{integralequation2}) for the 2-point function $S^{(2)}(t_1,t_0)$,
\beq\label{integralequation4}
S^{(3)}(t_1,t_0)=S^{(3)}_0(t_1,t_0)-\int_{t_0}^{t_1} {\rm d}t_3\,\int_{t_0}^{t_3} {\rm d}t_2\, S^{(3)}_0(t_1,t_3)\,\Sigma^{(3)}(t_3,t_2) \, S^{(3)}(t_2,t_0)\,.
\eeq
with ($X\equiv(\x,\y;\bar\x)$ and similarly for $X'$)
\beq\label{Sigma3a}
(X|\Sigma^{(3)}(t,t')|X')\equiv \delta(X-X')\delta(t-t')\,\frac{N_cn}{4}\left[\sigma(\x-\bar\x)+\sigma(\y-\x)+\sigma(\y-\bar\x)\right] ,
\eeq
where here  $\delta(X-X')=\delta(\x-\x')\delta(\y-\y')\delta(\bar\x-\bar\x')$.
The three dipole cross sections correspond to the three possible pairing of the lines in Fig.~\ref{3-point}.

In the calculation of the radiative corrections to the momentum broadening, the dominant logarithmic contribution to the 3-point function  involves  a very soft gluon,  carrying a small fraction $1-z\ll 1$ of the energy $E$ of the hard gluon. We call $\omega$ the energy of this soft gluon, i.e., 
\beq
 \omega\equiv(1-z)E\,.
\eeq
When the soft gluon  travels a distance $\Delta y_\perp \sim\sqrt{ 2\tau/\omega}$ during a given time $\tau$, the hard gluon travels a much smaller distance $\sqrt{2\tau/ E}\sim \sqrt{(1-z)}\Delta y_\perp$ (cf. Eq.~(\ref{freecalGx})). One may then approximate the 3-point function by replacing the propagators $\cal G$ and $\cal G^\dag$ of the hard gluon  by trivial Wilson-lines, but keeping the propagator of the soft gluon as a usual propagator. The resulting 3-point function   acquires then the following form 
\beq\label{3-point-tilde2}
(X_1|S^{(3)}|X_0)=\delta(\x_1-\x_0)\delta(\bar\x_1-\bar\x_0) \frac{1}{N_c(N_c^2-1)}\left<  (\y_1|{\cal G}^{a_1a_0}|\y_0){\rm Tr} \left( U^\dagger_{\bar \x_0}T^{a_1}U_{\x_0} T^{a_0} \right)   \right>,
\eeq
which we rewrite as
\beq
(X_1|S^{(3)}(t_1,t_0)|X_0)
=\delta(\x_0-\x_1)\delta(\bar\x_0-\bar\x_1)\,\tilde S^{(3)}(\u_0,\u_1,\v) , 
\eeq
with
\beq\label{3-point-tilde-xperp}
\tilde S^{(3)}(\u_0,\u_1,\v) &=&\int {\cal D}\u\exp\left\{\frac{i\omega}{2} \int_{t_0}^{t_1} \rmd t ~\dot{\u}^2 -\frac{N_c\, n}{4}\int_{t_0}^{t_1} \rmd t\,   \left[\sigma(\u)\!+\!\sigma(\v\!+\!\u)\!+\!\sigma(\v)\right]\right\}.\nn
\eeq
We have set $\v_0\equiv \x_0-\bar\x_0$, $\v_1\equiv \x_1-\bar\x_1$. The path integral in Eq.~(\ref{3-point-tilde-xperp}), which we shall refer to as the reduced 3-point function, 
depends non trivially only on $\v=\v_0=\v_1$, the (constant) size of the dipole representing the hard gluon, and on the end points of the path integral, namely $\u_1=\y_1-\x_1$ and $\u_0=\y_0-\x_0$.
In momentum space, we have, with $P_0\equiv(\p_0,\q_0;\bar \p_0)$ and $P_1\equiv(\p_1,\q_1;\bar \p_1)$,
\beq\label{eq:3-PF-FT}
 (P_1|S^{(3)}(t_1,t_0)|P_0)&=& (2\pi)^2\delta^{(2)}(\q_0\!+\!\p_0\!-\!\q_1\!-\!\p_1\!+\!\l) \theta(t_1-t_0)\,\tilde S^{(3)}(\q_0,\q_1,\l) \,,
 \eeq
where we have factored out the momentum conserving delta function and set $\l\equiv\bar\p_1-\bar\p_0$. It is easily seen that the  reduced 3-point function $\tilde S^{(3)}(\q_0,\q_1,\l)$ is given by
\beq\label{3-point-tilde}
 \tilde S^{(3)}(\q_0,\q_1,\l)&\equiv&  \! \int \rmd\u_0 \rmd\u_1 \rmd\v \;{\rm e}^{i\u_0\cdot \q_0-i\u_1\cdot\q_1-i\v\cdot\l} \,\tilde S^{(3)}(\u_0,\u_1,\v) . 
 \eeq
 It depends on only three momenta: the momentum $\p_0$ conjugate to the initial endpoint  $\u_0=\y_0-\x_0$ of the path integral over $\u$, the momentum $\p_1$ conjugate to the final endpoint $\u_1=\y_1-\x_1$ and the momentum $\l\equiv\bar\p_1-\bar\p_0$  conjugate to the (fixed) size $\v=\x-\bar\x$ of the  dipole associated with the hard gluon. In addition, $S^{(3)}$ depends on the times $t_0$ and $t_1$ (see Fig.~\ref{3-point}).

In the harmonic approximation, Eq.~(\ref{3-point-tilde}) reads  (it can be obtained for instance by taking the limit $z\to 1$ of Eq. ~(A.13) of Ref.~\cite{Blaizot:2013vha}),
\beq\label{S3harm}
&& \tilde S^{(3)}(\q_0,\q_1,\l)=\frac{16\pi }{3\hat q\tau}\exp\left\{-\frac{4
\big[\l-(\q_1-\q_0)/2\big]^2}{3\hat q\tau}\right\}\nn
&&\quad\times\ \frac{2\pi i}{\Omega\omega \sinh(\Omega\tau)}\exp\left\{-i\frac{(\q_0+\q_1)^2}{4\omega\Omega\coth(\Omega\tau/2)}-i\frac{(\q_1-\q_0)^2}{4 \omega\Omega\tanh(\Omega\tau/2)}\right\},
\eeq
where $\tau\equiv t_1-t_0$, and
\beq
\label{Omega}
\Omega\equiv\, \frac{1+i}{2\tau_{_{\rm br}}(\omega)},\eeq
with
$\tau_{_{\rm br}}(\omega)
\equiv \sqrt{{\omega}/{\hat q}}$.
This expression will be used later in the evaluation of the radiative correction to $\hat q$.

\subsection{Single radiative correction}

\begin{figure}[htbp]
\begin{center}
\includegraphics[width=5cm]{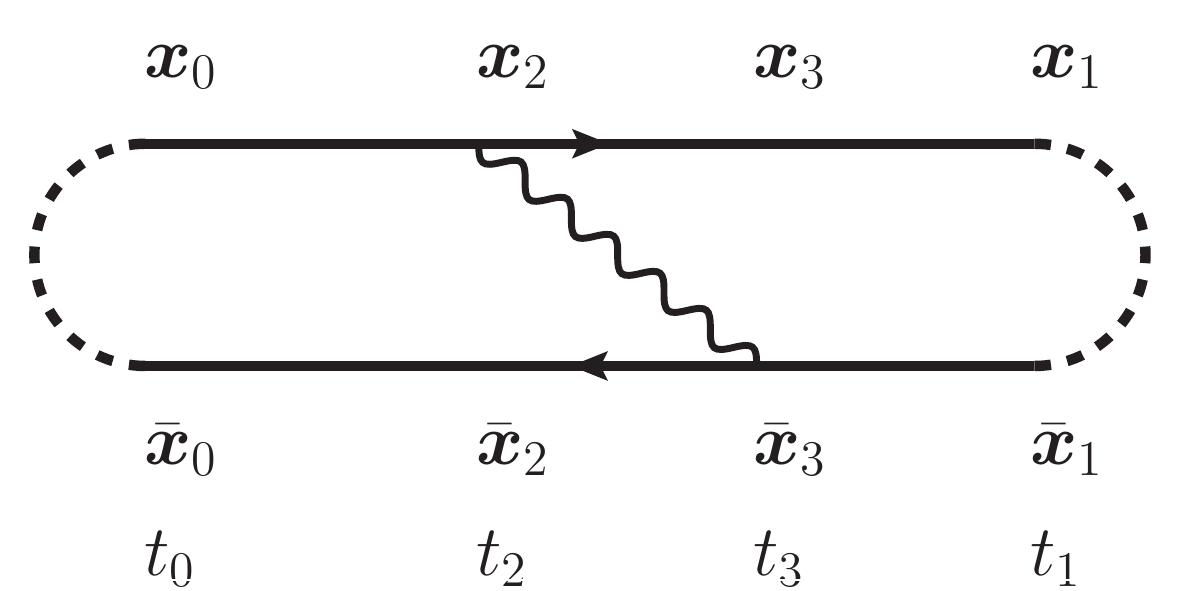}\qquad$+$\qquad\includegraphics[width=5cm]{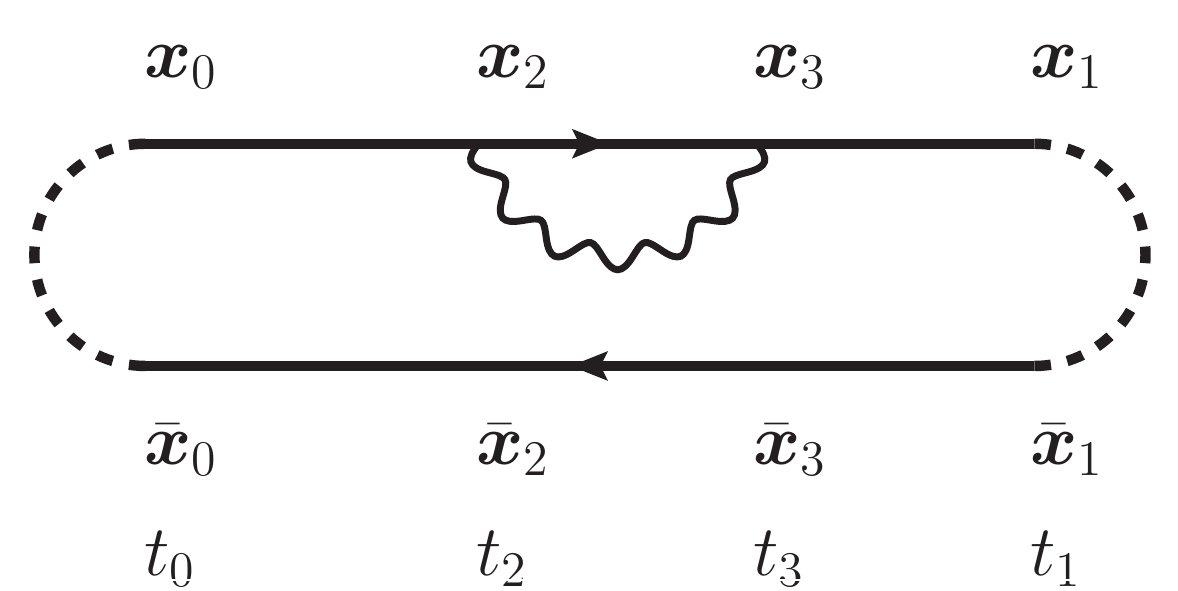}\\

\caption{Real (left) and virtual (right) contributions to the radiative correction to the 2-point function. There are other diagrams that are not shown, corresponding to a different time ordering of the emission and absorption vertices in the real term, and a diagram where the virtual correction is attached to the propagator ${\cal G}\dagger$, i.e., to the lower line. The contributions of these extra diagrams can be accounted for by taking twice the real part of the diagrams shown (see Appendix). }
\label{fig5}
\end{center}
\end{figure}

We now consider explicitly the radiative corrections displayed in Fig.~\ref{fig5}.  These diagrams represent a correction to the 2-point function, with, as in the previous section, the amplitude and its complex conjugate drawn one above each other: the upper thick line is  a propagator ${\cal G}$ in the amplitude of the hard gluon, the lower thick line  a propagator ${\cal G}^\dagger$ in the complex conjugate amplitude. The wavy line represents the soft radiated gluon of energy $\omega$. Strictly speaking, the corresponding  propagator is a time-ordered propagator, but  this plays no role in the present calculation. 

The modification of the probability  ${\cal P}$ due to a radiative correction  is given by Eq.~(\ref{deltaP}) in the Appendix, which involves a 3-point function and ``external'' factors ${\cal P}$. Noticing that  the correction that we are looking for is contained in the reduced 3-point function $ \tilde S^{(3)}$ and {\em not} in the external factors ${\cal P}$, we replace these external factors by their free values (given for instance by the first term in Eq.~(\ref{ME})). One gets then, after relabeling the momenta,
 \beq\label{deltaP}
\Delta {\cal P}(\p_1\, t_1|\p_0\, t_0)&=&\frac{g^2N_c}{4\pi}\,2\text{Re} \,\int
\frac{\rmd \omega}{\omega^3}\,\int_{t_0}^{t_1}\rmd t_3\int_{t_0}^{t_3}\rmd t_2 \nn
 &\times&\int_{ \q_2\q_3} (\q_2\cdot \q_3) \left[ \tilde S^{(3)}(\q_2, \q_3,\l+\q_3) -\tilde S^{(3)}(\q_2, \q_3,\l)  \right],\nn
\eeq
with here $\l=\p_1-\p_0$, and both reduced 3-point functions depend on $t_3-t_2$.  One recognizes in this expression the contributions of the two diagrams displayed in Fig.~\ref{fig5}. Note that 
\beq\label{deltaPproba}
\int_{\p_1}\Delta{\cal P}(\p_1\, t_1|\p_0\, t_0)=0,
\eeq
as it should be for $\Delta{\cal P}$ to be interpreted as a correction to a probability.

 At this point, we rewrite the time integrals as follows
\beq
 \int _{t_0}^{t_1}d t_3 \int _{t_0}^{ t_3}dt_2 =  \int _{t_0}^{t_1}d t_2 \int ^{t_1-t_2}_{0}d\tau 
 \eeq
 with $\tau=t_3-t_2$. We then make a  ``local'' approximation, i.e., we assume that  the double integral is dominated by values of $\tau$ that are  much smaller than the typical intervals covered by either  $ t_2$ or $t_3$ (which are  of order $L$). This allows us to factor out the integration over $\tau$, and to treat the radiative correction as a local correction. We get then, after performing the $t_2$ integration, which yields a factor $t_1-t_0$, 
  \beq\label{deltaP2}
&&\frac{\Delta {\cal P}(\p_1\, t_1|\p_0\, t_0)}{t_1-t_0}=\nn &&\qquad\qquad\frac{g^2N_c}{4\pi} \,2\text{Re} \,\int
\frac{\rmd \omega}{\omega^3}\,\int\rmd \tau \int_{ \q_2\q_3} (\q_2\cdot \q_3) \left[ \tilde S^{(3)}(\q_2, \q_3,\l+\q_3;\tau) -\tilde S^{(3)}(\q_2, \q_3,\l;\tau)  \right],\nn
\eeq
where we have left the boundaries of the $\tau$ integration unspecified. These will be dealt with shortly. 
In fact, there is a subtlety here. As we shall see, the integral over $\tau$ is divergent at small $\tau$, but only logarithmically, so that  it is not really dominated by the  smallest allowed values of $\tau$.  However, as we shall argue later, the upper end of the integration does not affect the coefficient of the logarithm, which is all what we can calculate at this stage.    
 
 From Eq.~(\ref{deltaP2}) it is a simple matter to identify the transport coefficient. To do so, in line with the remark above concerning the effective locality of the radiative correction,  we assume that the broadening distribution obeys, in the presence of the radiative correction,  an equation similar to Eq.~(\ref{ME}), with a modified dipole cross section. We then isolate the change in the distribution, $\Delta{\cal P}(\p_1\, t_1|\p_0\, t_0)$,  associated to one radiative correction occurring between $t_0$ and $t_1$,  by considering a small interval $t_1-t_0$, setting  ${\cal P}={\cal P}_0$ in the right hand side of Eq.~(\ref{ME}), and absorbing  the entire modification into a change of the dipole cross section that we denote by $\Delta\sigma$. We get then
 \beq\label{Deltasigma}
 \frac{\Delta{\cal P}(\p_1\, t_1|\p_0\, t_0)}{t_1-t_0}=-\frac{N_c}{2}n \Delta\sigma(\p_1-\p_0)\,.
 \eeq
 By comparing Eq.~(\ref{Deltasigma}) and Eq.~(\ref{deltaP2}), one obtains
 \beq
-\frac{N_c}{2}n \Delta\sigma(\l)=\frac{g^2N_c}{4\pi} \,2\text{Re} \,\int
\frac{\rmd \omega}{\omega^3}\,\int\rmd \tau \int_{ \q\q'} (\q\cdot \q') \left[ \tilde S^{(3)}(\q, \q',\l+\q';\tau) -\tilde S^{(3)}(\q, \q',\l;\tau)  \right],
\eeq
 and, referring to Eq.~(\ref{qhat0}), one deduces the correction to the jet-quenching parameter
 \beq
\Delta \hat q\,=\alpha_s N_c \,2\text{Re}\int \frac{\rmd \omega}{\omega^3}\,\int d\tau\int_{\q,\q',\l} \,[(\q'-\l)^2-\l^2]\, (\q\cdot\q')\,\tilde S^{(3)} (\q,\q',\l;\tau)\,,
\eeq
where we have set $\alpha_s=g^2/(4\pi)$. Note that, in contrast to what happens for the case of instantaneous interactions with the medium where only the real term contributes (see Eq.~(\ref{qhat0}) and the remark below it), here both the real and the virtual terms contribute to the correction $\Delta\hat q$. This is because at least one interaction with the medium occurs during the lifetime of the fluctuation, and this changes the transverse momentum. This change is reflected in both the real and the virtual contributions. 

It is instructive to see how  the same correction can be obtained by carrying out the calculation in coordinate space. This provides an interesting and complementary view, that besides will be useful for the discussion in the next section, where the use of coordinate space will prove useful in handling the radiative corrections to the 3-point function, in particular for the treatment of the color algebra. 
In coordinate space,  it is  the 2-point function, $\Delta \SII (t_1,t_0)$ that is the appropriate object to consider, rather than  the momentum broadening probability itself. The calculation involves the same approximations as those used in momentum space. In particular the coordinates of the hard gluons (in the amplitude and its complex conjugate) are frozen during the lifetime of the fluctuation, i.e., between $t_2$ and $t_3$,  and accordingly, the hard gluon propagators are given by trivial Wilson lines (cf. Eq.~(\ref{path-int2})),
\beq\label{prop-E}
(\x_3|{\cal G}|\x_2)\approx \delta(\x_3-\x_2)\,U_{\x_2}\,\quad\text{and}\quad(\bar\x_2| {\cal G}^\dag|\bar\x_3)\approx \delta(\bar\x_3-\bar\x_2)\,U^\dag_{\x_2}.
\eeq
The calculation leads then, for the real correction, and using the matrix notation of Eq.~(\ref{integralequation2}), 
\beq\label{DS2-2a}
&&\Delta \SII (t_1,t_0)=  -\int _{t_0}^{t_1}d t_3 \int _{t_0}^{t_3}dt_2  \,  \SII (t_1,t_3)\,\Delta \Sigma^{(2)}(t_3,t_2)\,\SII (t_2,t_0)\,, 
\eeq
with 
\beq\label{DS2-2b}
&&(X_3|\Delta\Sigma^{(2)}(t_3,t_2)|X_2) \nn
&&=\delta(X_3-X_2)\frac{\alpha_s}{N_c^2-1}\,2\text{Re}\int \frac{d\omega}{\omega^3} \left\langle \text{Tr}\,U^{\dag}_{\bar\x_2} T^{a_3}U_{\x_2} T^{a_2}\,\left[\bdel_{\y_2}\cdot\bdel_{\y_3}\,(\y_3|\Gc^{a_3 a_2}|\y_2)\right]_{\y_2=\x_2,\y_3=\bar\x_2}\right\rangle\nn
&&=\delta(X_3-X_2)\frac{\alpha_sN_c}{2}\,2\text{Re}\int \frac{d\omega}{\omega^3}\,\bdel_{\u_2}\cdot\bdel_{\u_3}\tiSIII(\u_2,\u_3,\v;\tau)|_{\u_2=0, \u_3=-\v}\,,\nn
\eeq
where $\tau\equiv t_3-t_2$, $\u_2\equiv\y_2-\x_2 $, $\u_3\equiv \y_3-\x_3 $ and $\v\equiv \x_2-\bar\x_2$, and we have used  the eikonal vertex acting on the the radiated gluon propagator 
\beq\label{eikonalGamma2}
{\bs\Gamma}_\x^a=\frac{2ig}{z} T^a  i\bdel_{\x}\,,\qquad (T^a)_{bc}=if^{abc}\,,\qquad z=\omega/E\,.
\eeq
  The virtual correction is given by the same formula in which $\bdel_{\u_2}\cdot\bdel_{\u_3}\tiSIII(\u_2,\u_3,\v;\tau)|_{\u_2=0,\u_3=-\v}$ is replaced by $-\bdel_{\u_2}\cdot\bdel_{\u_3}\tiSIII(\u_2,\u_3,\v;\tau)|_{\u_2=\u_3=0}$. \\

By the same reasoning as before, we shall treat the correction $\Sigma^{(2)}(t_3,t_2)$ as a local correction, i.e., set formally $\Sigma^{(2)}(t_3,t_2)\propto \delta(t_3-t_2)$.
 After replacing the factors $S^{(2)}$ external to the radiative correction by free 2-point functions (which are independent of time), and performing the $t_2$ integration, we get
\beq\label{DS2-3}
&&(X_1|\Delta \SII (t_1,t_0)|X_0)= -(t_1-t_0)\delta(X_1-X_0)\frac{\alpha_sN_c}{2}\,\int \frac{d\omega}{\omega^3} \int  d\tau  K(\v,\tau),
\eeq
with
\beq\label{K-x}
K(\v,\tau)&\equiv &2\text{Re}\,\bdel_{\u_2}\cdot\bdel_{\u_3}\left[\tiSIII(\u_2,\u_3,\v;\tau)|_{\u_2=\0, \u_3=-\v} -\tiSIII(\u_2,\u_3,\v;\tau)|_{\u_2= \u_3=\0}\right]\,,\nn
&=& 2\text{Re} \int_{\q_2, \q_3, \l}\left( e^{i(\l-\q_3)\cdot\v}-e^{i\l\cdot\v}\right)\,
(\q_2\cdot\q_3)\tiSIII(\q_2,\q_3,\l;\tau).
\eeq
Note the vanishing of this expression when $\v\to 0$. This is the equivalent, in the coordinate space representation, of the relation (\ref{deltaPproba}) expressing the fact that $\Delta{\cal P}$ is a probability. It results from a cancellation between real and virtual terms, reflecting the property of color transparency of a dipole cross section. The identification of the correction to the dipole cross section proceeds this time by comparison with the integral equation (\ref{integralequation2}) (more properly, its generalization, whose validity, as that of the Fokker-Planck equation relies on the  locality assumption, as emphasized earlier), as well as Eq.~(\ref{sigmaSigma2}) for fixing the relation between $\Delta \sigma$ and $\Delta\Sigma^{(2)}$. One then gets, leaving the bounds on the $\tau$ integration unspecified for the moment, 
\beq\label{dsig-1}
\frac{N_c n}{2} \Delta\sigma(\v)&=&\alpha_s \frac{N_c }{2} \int \frac{d\omega}{\omega^3} \int d\tau\,K(\v,\tau)\approx \frac{1}{4}\v^2\Delta\hat q,\nn
\eeq
where the last equality stems from Eq.~(\ref{qhat-r2}) expressing $\sigma(\v)$ at small $\v$ in the harmonic approximation. We use this relation to interpret the correction $\Delta\sigma(\v)$ as a correction to the parameter $\hat q$. 

We now move to the explicit calculation of $\Delta\hat q$. Using the explicit expression (\ref{S3harm}) of the reduced 3-point function, one can easily perform the integrations over transverse momenta and get 
\beq\label{dsig-0}
\Delta \hat q\,&=&2\alpha_s N_c  \int \frac{d\omega}{\omega^3} \int d\tau \,2\text{Re} \int_{\q,\q', \l}\frac{1}{2}\left[(\q'-\l)^2-\l^2\right]\,
(\q\cdot\q')\tiSIII(\q,\q',\l;\tau)\,,\nn
&=&  \, \frac{\alpha_s \,N_c}{\pi}\, 2\text{Re }
\int d\omega \,\int \rmd \tau  \,\frac{i \Omega^3}{\sinh(\Omega \tau)}\left[1+\frac{4}{\sinh^2(\Omega  \tau)}\right]\,,\nn
&\simeq&   \, \frac{\alpha_s \,N_c}{\pi}\, 2\text{Re }
\int d\omega \,\int \rmd \tau  \,\frac{i \Omega^3}{\Omega \tau}\,,\nn
&\simeq & \frac{\alpha_s \,N_c }{\pi}\, \,
\int  \frac{\rmd\omega}{\omega}\int
\frac{\rmd\tau}{\tau}\,\hat q\,,\nn
\eeq
which exhibits a double logarithmic divergence when $\tau\to 0$.
The boundaries of the integrations have been discussed extensively already \cite{Liou:2013qya,Blaizot:2013vha}.  As clear from Eq.~(\ref{dsig-0}) the $\tau$ integral is bounded at the upper end but $\tau_{\rm br}(\omega)=\sqrt{\omega/\hat q}$ corresponding to the onset of the multiple scattering regime and the LPM effect (see Eq.~(\ref{S3harm})). In order to specify more completely the boundaries of the double integral, it is more convenient to change variables, from $(\omega,\tau)$ to $ (\q, \tau)$,  
with $\tau\equiv \omega/\q^2$ the formation time of the radiated gluon, and $\q$ its transverse momentum which can run up to $\p^2\equiv (\p_1-\p_0)^2$. We obtain then
\beq\label{qhat1}
\Delta \hat q (\tau_\text{max}, \p^2)\equiv \frac{\alpha_sN_c}{\pi} \, \int^{\tau_\text{max}}_{\tau_0} \frac{\rmd\tau}{\tau}\int^{\p^2}_{\hat q\tau}\frac{\rmd\q^2}{\q^2}\, \hat q (\q^2)\,,
\eeq
where we have explicitly indicated the scale dependence of $\hat q$. The boundary corresponding to the region of multiple scattering now appears as the lower bound of the $\q$ integration,  $ \q < k_\text{br}\equiv \hat q\tau $.  The largest value of $\tau$ is now $\tau_{\rm max}\sim t_1-t_0\sim L$, while $\tau_0^{-1} $ may be viewed as the largest energy that can be extracted from the medium through a single scattering.  For a constant $\hat q $ in the integral, one can easily perform the integrations, and by keeping the leading contributions, we recover the result first derived in \cite{Liou:2013qya,Blaizot:2013vha}, 
\bal \label{deltaqhat}
 \Delta \hat q (\p^2)\simeq  \frac{ \alpha_s C_A}{2\pi}\, \hat q \ln^2\left(\frac{\p^2}{\hat q \tau_0}\right).
\end{align}

\subsection{Independent multiple radiative corrections }\label{multiple-rad}

We return now to the approximation that we have used in order to identify the radiative correction to the 2-point function as a correction to the jet quenching parameter. The main assumption is the locality in time. We know however that this is not a priori a good approximation since the radiated soft gluons can have  long lifetimes, and successive radiations can overlap, ruining the reasoning based on the Fokker Planck equation, or the exponentiation of the 2-point function. Let us recall that in the absence of radiative corrections, the  factorization of successive interactions with the medium particles relies on the instantaneity of these interactions, so that successive collisions do not overlap, and also on the fact that the color structure is trivial, so that the 2-point function remains a singlet after each interaction. These  two properties no longer hold in the presence of radiative corrections. 
Still we shall now argue that the logarithmic nature of the phase space integration  allows us, in fact, to treat the dominant corrections as if they were effectively local. This is so because uncertainties in the upper limit of the logarithmic integration leads only to sub-leading contributions.

This argument can be best illustrated by a simple calculation.  We focus on the time integrations, since the phase space is the most important part here (it is indeed unlikely that peculiar color structures of overlapping radiations amplify a singularity). The exponentiation of the interactions result from the simple property of the integral giving for instance the survival probability of a particle propagating from $t_0=0$ to $L$, and undergoing independent successive collisions with cross section $\sigma_0$:
\beq\label{S0}
S(L)=\sum_{n=0}^\infty \int_0^L \rmd t_n\int_0^{t_n} \rmd t_{n-1} ...\int^{t_2}_0 \rmd t_1 (-\sigma_0)^n \equiv \exp\left(-\sigma_0 L\right)\,.
\eeq
Let us now consider a radiative correction leading to a logarithmic contribution of the form
\beq\label{model-rad1}
\Delta S(L)=-\sigma_0\,\alpha\int_0^L \rmd t \int_{\tau_0}^{L} \frac{\rmd\tau}{\tau}\equiv \alpha\,\ln\frac{L}{\tau_0}\,(-\sigma_0L)\,.
\eeq
\begin{figure}[htbp]
\centering
\includegraphics[width=7cm]{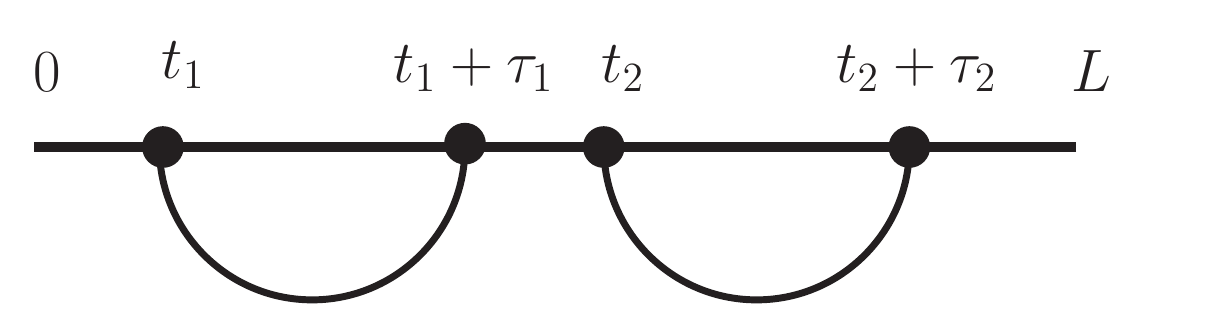} \qquad\includegraphics[width=7 cm]{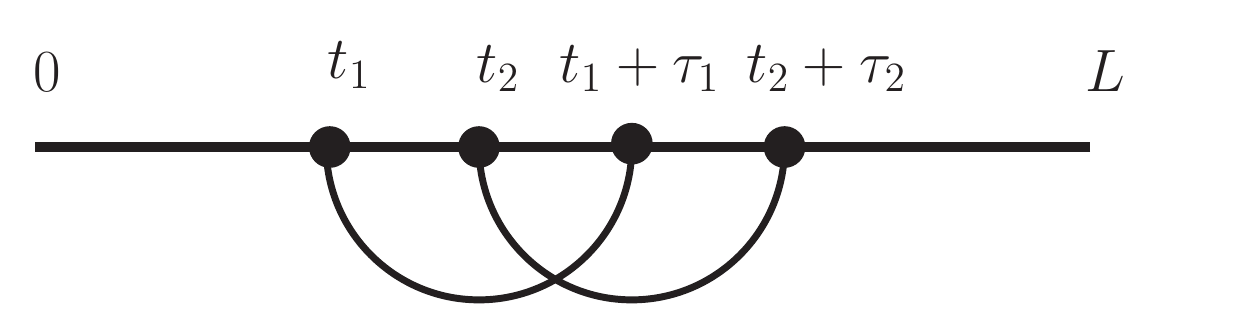} \\
\caption{Two radiative contributions corresponding to different time integration regions. Left: disconnected graph. Right: connected graph. }\label{fig:examples}
\end{figure}
The processes with two radiative corrections correspond to the two graphs displayed in Fig.~\ref{fig:examples}. The disconnected graph (left graph in Fig.~\ref{fig:examples}),  yields, when $\alpha \ln(L/\tau_0)\sim1$ and $\alpha\ll1$, the contribution
\beq
(\alpha\,\sigma_0)^2\int_{\tau_0}^L \rmd t_2 \int_0^{t_2-\tau_0} \rmd t_1 \int_{\tau_0}^{L-t_2}\frac{\rmd \tau_2}{\tau_2}\int_{\tau_0}^{t_2-t_1}\frac{\rmd \tau_1}{\tau_1}\simeq \frac{1}{2!} \left(-\sigma_0L\, \alpha\,\ln\frac{L}{\tau_0}\right)^2\,,
\eeq
while the connected graph  yields a contribution $\sim (\alpha \sigma_0 L)^2 \sim \,{\cal O}(\alpha^2)$,
which is  suppressed relative to the contribution of the disconnected graph (it does not contain a logarithmic enhancement). 
Thus, even though the two fluctuations overlap, this is of no consequence for the leading logarithmic corrections; this plays no role in  particular in the determination of the coefficient in front of the logarithmic correction. It follows therefore that the successive logarithmic corrections can be considered as effectively independent, allowing for the exponentiation of the disconnected graphs
\beq
S(L)=\exp\left[-\sigma_0\left(1+\alpha\ln\frac{L}{\tau_0}\right)\right] \simeq S_\text{disc}(L)+ {\cal O}(\alpha)\,.
\eeq
It can furthermore be verified that the disconnected graphs  remain the leading (logarithmic) contributions when we allow for mixed contributions involving the instantaneous interactions that lead to (\ref{S0}).

This argument is what allows us to generalize Eq.~(\ref{integralequation2}) to a $\Sigma(t,t')$ that includes the radiative corrections, to treat the double logarithmic correction to  $\Sigma(t,t')$ as   effectively local, and multiple such radiative corrections as independent. As we have seen, such corrections can be interpreted, in the coordinate space description, as a modification of the dipole cross-section $
\sigma(\v)\,\to\, \sigma(\v)+\Delta\sigma(\v,\tau_\text{max})$, 
where $\tau_\text{max}\equiv t_1-t_0$, or in momentum space as a modification of the equation for the momentum broadening probability 
\beq\label{P-mom-FP-rad}
\frac{\del}{\del t}{\cal P}(\p,t) = \frac{1}{4}\,\left(\frac{\del}{\del \p}\right)^2\,\left[ \hat q(\p^2)+\Delta\hat q(t,\p^2)\right]\,{\cal P}(\p,t)\,,
\eeq
which generalizes Eq.~(\ref{P-mom-FP}). In both cases, the radiative corrections are accounted for by a correction to the jet quenching parameter.  

\section{Radiative corrections to the medium-induced gluon spectrum}\label{rad-E-loss}

We now address the main issue that motivated this paper, namely how the correction to $\hat q$ that we have identified in the previous section affects the spectrum of radiated gluons, and more generally the medium-induced branching processes that give rise to the in-medium QCD cascade. We shall present a general analysis for the former case, while the general branching process will be dealt with only in the limit of large $N_c$.

 Let us recall  that the spectrum (\ref{BDMPS}) of radiated gluons with frequency  $\omega\ll E$, can be calculated from the reduced 3-point function (\ref{3-point-tilde}), according to
  \beq\label{kernel5}
\frac{\rmd N}{\rmd\omega \rmd t}\equiv\, \frac{\alpha_sN_c}{\omega^2}\, 2\text{Re}
\int_{0}^{\infty} \rmd\tau\int_{\q,\q',\l } (\q\cdot\q')\, \tilde S^{(3)}(\q,\q', \l;\tau)\,.
\eeq
Here the variable $t$ runs up to $\sim L$. This spectrum is valid in the large medium length limit where the gluon branching time $\tau_{\rm br}=\sqrt{\omega/\hat q}\ll L$ and the integration over the gluon formation time $\tau < L$ is suppressed exponentially beyond $\tau_{\rm br}$. This is why we can integrate $\tau$ up to infinity in Eq.~(\ref{kernel5}).

The 3-point function $\tilde S^{(3)}$ is given explicitly by Eq.~(\ref{S3harm}) in the harmonic approximation, and $\tau=t_1-t_0$, with $t_0$ and $t_1$ denoting the times of the emission in the amplitude and its complex conjugate, respectively (see Fig. \ref{3-point}). The BDMPS spectrum of Eq.~(\ref{BDMPS}) is easily recovered from this expression by performing the integrations over the transverse momenta and over $\tau$ \cite{Mehtar-Tani:2013pia,Blaizot:2012fh}. As is clear from Eq.~(\ref{S3harm}) the reduced 3-point function depends explicitly on $\hat q$. Our goal is to show that the leading radiative corrections do not modify $\tilde S^{(3)}$, except for a  change in the value of the parameter $\hat q$, the correction to $\hat q$ being,  besides,  the same as that calculated in the previous section for momentum broadening.

Quite generally, we shall be concerned with the radiative corrections of the 3-point function (\ref{3-pt-def}), whose graphical interpretation is given in Fig.~\ref{3-point}.  The  diagram displayed there corresponds typically to a branching process where a gluon with initial energy $E$ (represented by the lower two thick lines in the amplitude and the conjugate amplitude, respectively), splits into two gluons with energies $zE$ and $\omega\equiv (1-z)E$, the latest being represented by the upper thick line. 
\begin{figure}[htbp]
\begin{center}
\includegraphics[width=6cm]{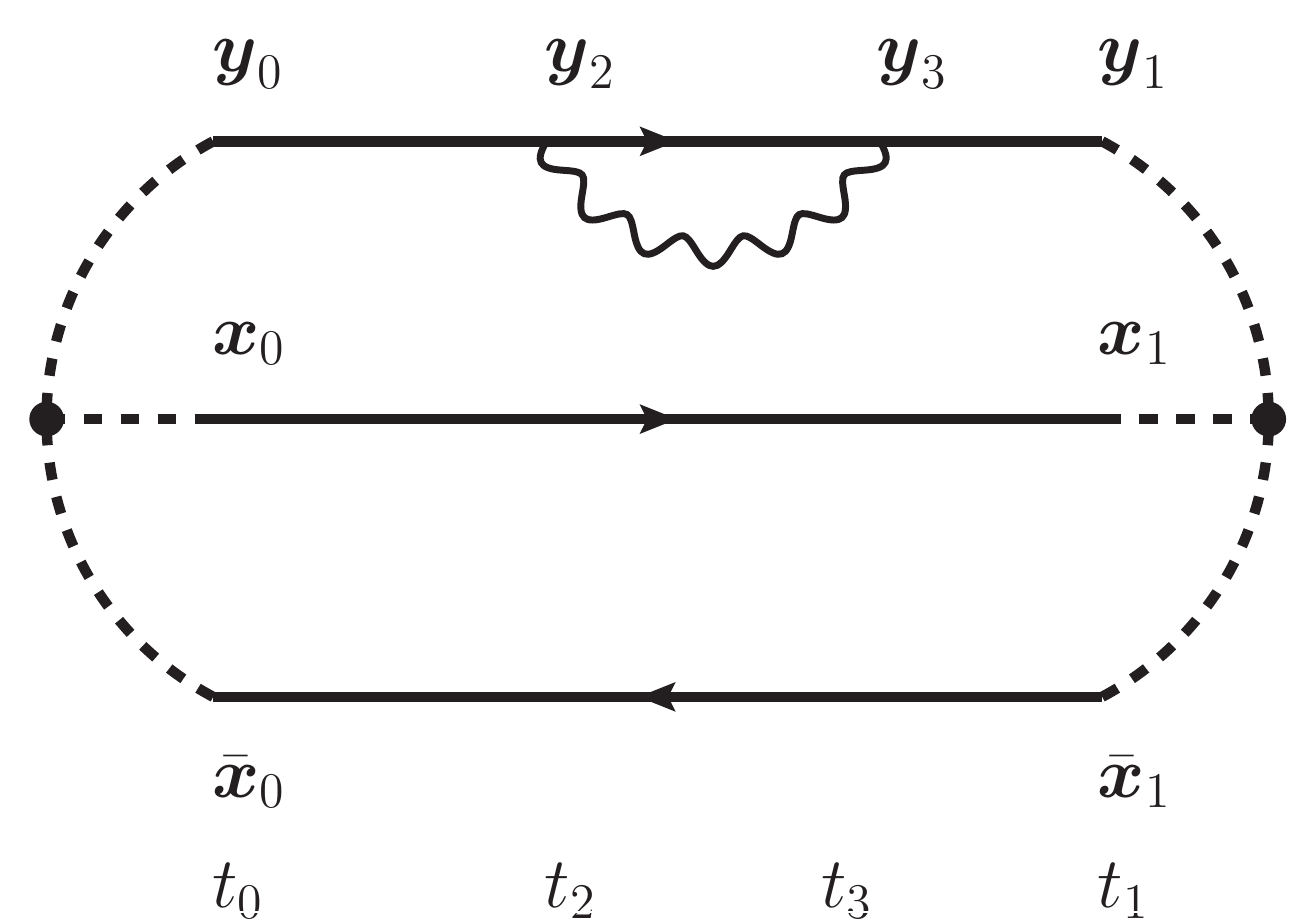} \includegraphics[width=6cm]{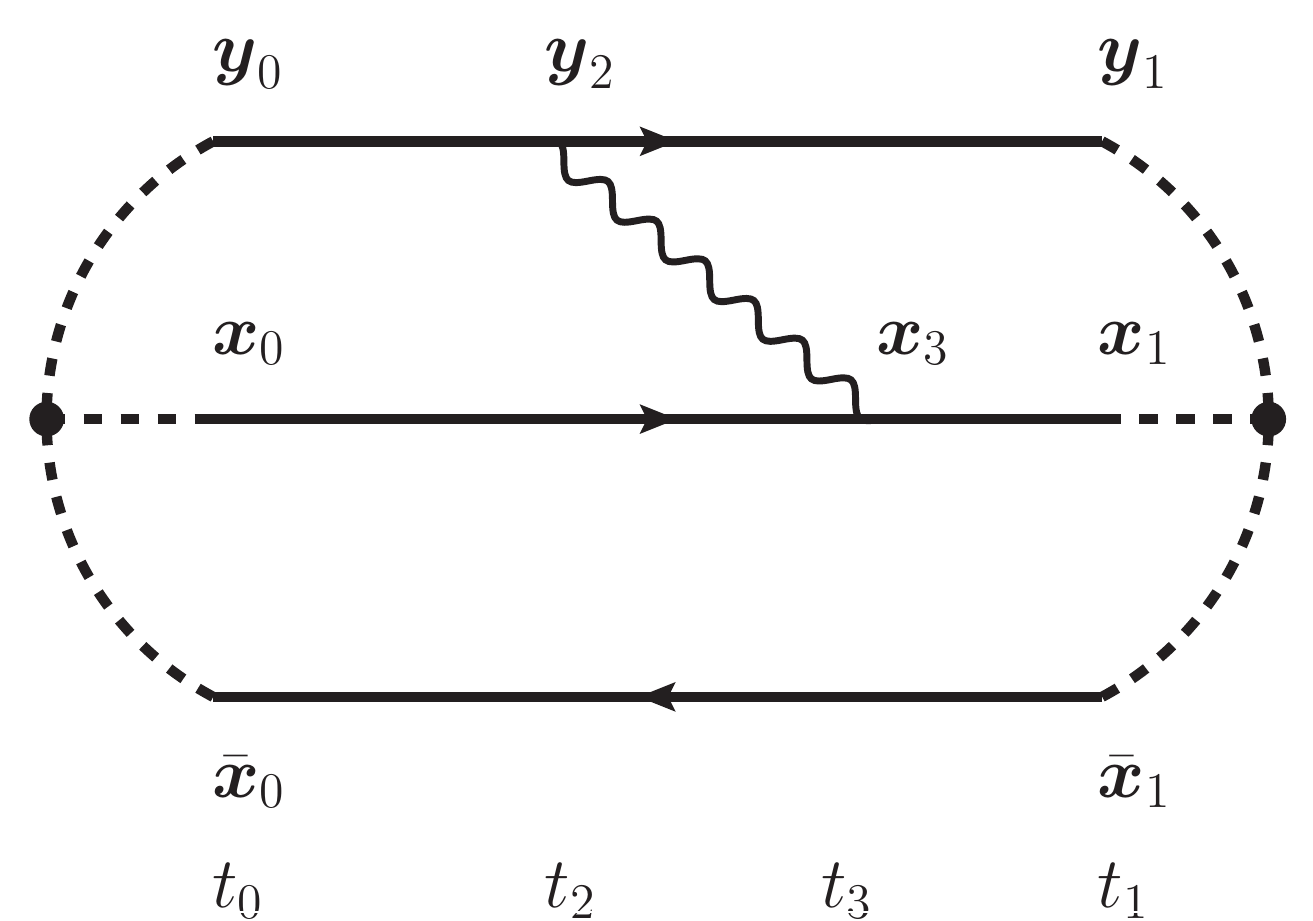}\includegraphics[width=6cm]{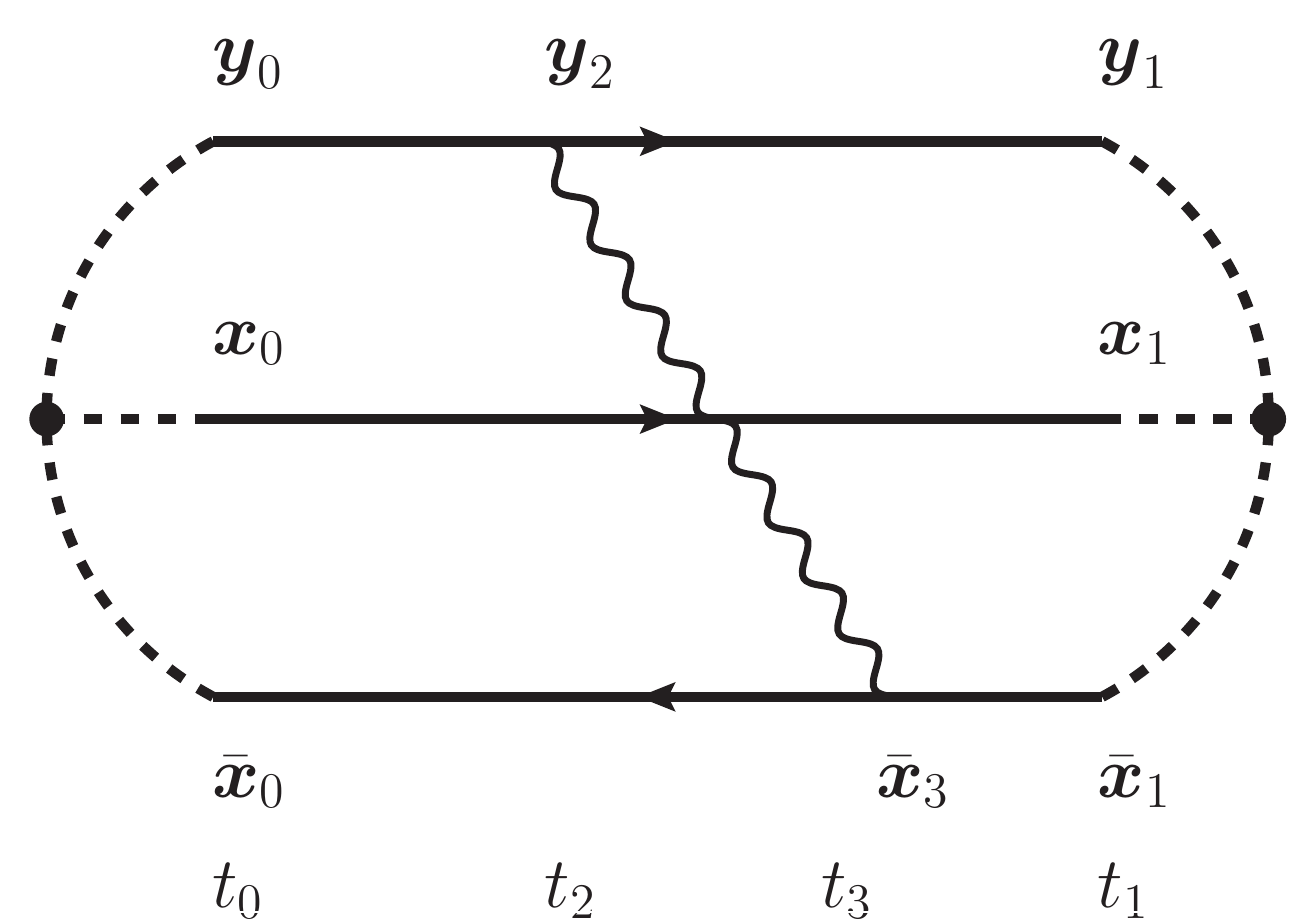}\\
\caption{Contributions $A_1$ (top), $A_2$ (bottom left)  and $A_3$ (bottom right) to radiative corrections of the 3-point function corresponding to radiation off the upper line (the gluon with energy $\omega=(1-z)E$). The lowest two lines represent the gluon, with energy $zE$, in the amplitude (middle line), and that with  energy $E$ in the complex conjugate amplitude (bottom line). In most of the discussion of this section, $z\simeq 1$, so that the gluon $\omega$ is soft, while the gluon $E$ is hard. }
\label{rad-3p}
\end{center}
\end{figure}

Diagrams representing the radiative corrections are displayed in Fig. \ref{rad-3p}. The gluon that is responsible of the radiative correction  carries energy $\omega'$ and is represented by a wavy line.  Altogether there are nine diagrams to consider, but only three of those are shown  in Fig.~\ref{rad-3p}. They correspond to the emission from the line $\y$ (gluon $\omega$) at time $t_2$ and ending at time $t_3$ either on the same line, or on the $\x$  and $\bar\x$  lines corresponding to the  gluon $E$. These  three contributions will be referred to as $A_1$, $A_2$ and $A_3$ respectively. Similarly, we have three diagrams $B_i$ and three $C_i$, with $i=1,2,3$, for emissions from the $\x$ and $\bar\x$ lines respectively. 

We expect  a double logarithmic enhancement when the  energy $\omega'\ll \omega,E$ of the radiated gluon becomes very small, or equivalently when the corresponding fluctuation is very short-lived:  $\tau'=t_3-t_2\ll\tau$, where the gluon $\omega'$ is emitted at time $t_2$ and reabsorbed at time $t_3$, while $\tau=t_1-t_0$ is the typical time scale of the branching process $E\to (zE,(1-z)E)$. Accordingly, and as we did earlier when calculating the radiative corrections to the 2-point function, we shall replace, in the time interval $[t_2,t_3]$, the propagators attached to the thick lines in the diagrams of Fig.~\ref{rad-3p} by trivial Wilson lines (see Eq.~(\ref{path-int2})), while that of the wavy line will be an ordinary propagator ${\cal G}$ (Eq.~(\ref{path-int})), or ${\cal G}^\dagger$ if the emission occurs on the line $\bar \x$. The vertex coupling the radiated gluon to the thick lines will be taken in the eikonal approximation and is given by Eq.~(\ref{eikonalGamma2}). 

\subsection{Correction to the gluon spectrum}

 For the calculation to the correction to the gluon spectrum,  we assume $z\lesssim1$, so that $\omega\ll E$. Furthermore, the  energy $\omega'$ of the radiated gluon is much smaller that the energy of its emitter, so that $\omega'\ll \omega\ll E$. To summarize, we look now at how the process where a gluon of energy $E$ splits into two gluons, one with energy $\omega\ll E$, is affected by the emission of an additional very soft gluon with energy $\omega'\ll \omega$. Note that in the diagrams shown in  in Fig. \ref{rad-3p}, the lowest two lines represent the hard gluon, with energy $\sim E$ (in the amplitude and complex conjugate amplitude), while the upper line represents the soft gluon with energy $\omega\ll E$. 

\begin{figure}[htbp]
\begin{center}
\includegraphics[width=10cm]{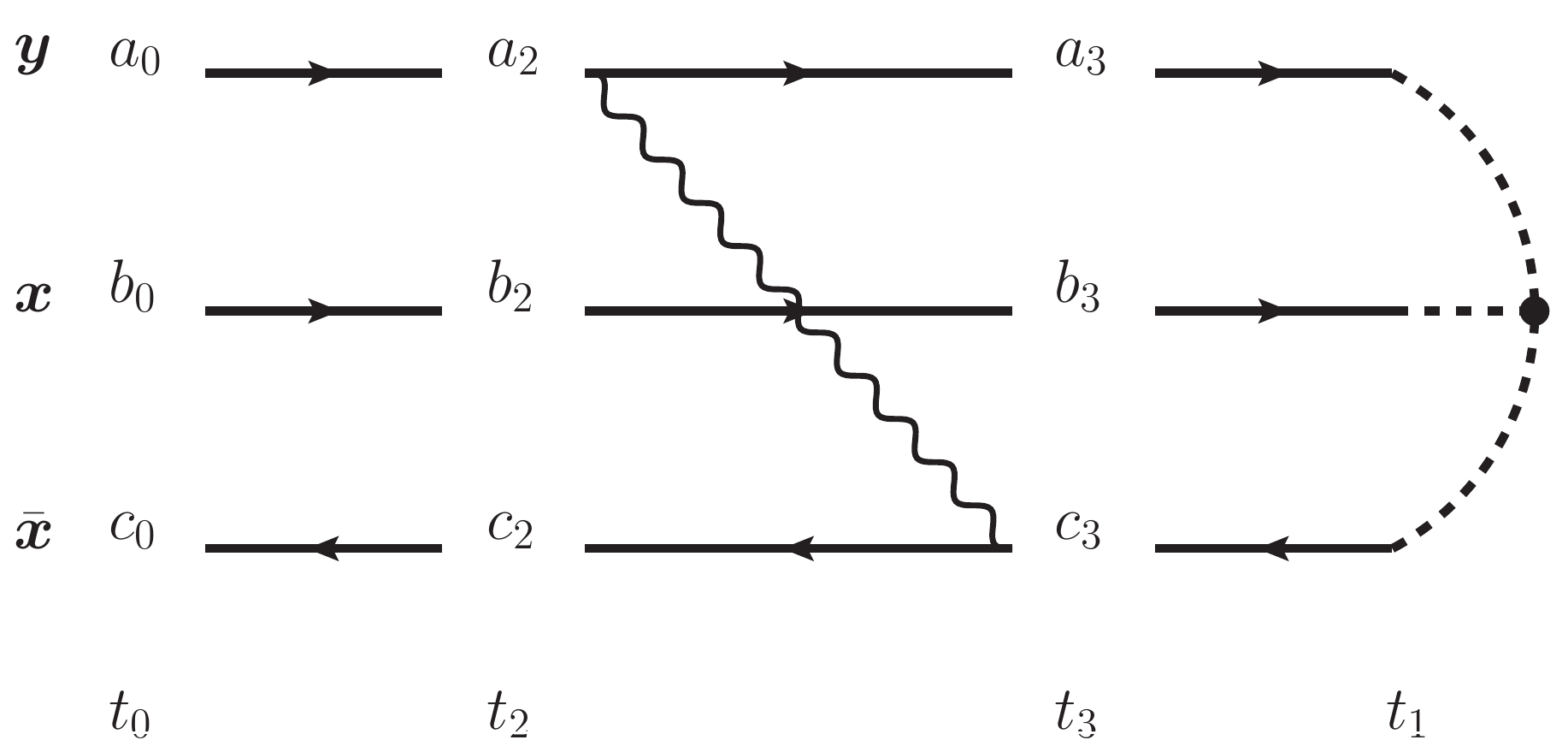}
\caption{Illustration of Eq.~(\ref{DS3}). This diagram represents the contribution $(A)_3$ to the tensor $F$,  Eq.~(\ref{DS3-A3}). Various color tensors appear in the three regions $[t_0,t_2]$, $[t_2,t_3]$ and $[t_3,t_1]$, respectively $\SIII_{a_2b_2c_2,a_0b_0c_0}$ from $t_0$ to $t_2$, $F_{a_3b_3c_3,a_2b_2c_2} $ from $t_2$ to $t_3$ and  $\SIII_{a_3b_3c_3}$ from $t_3$ to $t_1$.  The soft gluon  with frequency $\omega'$ (wavy line) is emitted at time $t_2$, and transverse coordinate $\y_2$ and reabsorbed at time $t_3$, at transverse coordinate $\bar\x_3$. Its non-eikonal propagation is described by the propagator $(\bar\x_3|{\cal G}(t_3,t_2)|\y_2)$. The three thick lines represent the eikonal propagation, described by trivial  Wilson lines, of gluon $\omega$ (the upper line) in the amplitude, and gluon $E$, the middle and the lower lines, in the amplitude and complexe conjugate amplitude respectively.  }
\label{rad-3p-A3}
\end{center}
\end{figure}

In Sect.~\ref{multiple-rad} we have argued that, to logarithmic accuracy, multiple radiative corrections can be treated as independent and ordered in time,  similarly to the instantaneous interaction approximation that yields Eq.~(\ref{integralequation4}). Therefore,  in analogy with what we did for the 2-point function in the previous section, we assume that the modified 3-point function obeys a generalization of Eq.~(\ref{integralequation4}), where the radiative correction plays the role of a modified dipole cross-section, and we isolate the contribution with one radiative correction\footnote{For reasons discussed at the end of the previous section, we exclude from the present analysis the case where successive radiative corrections overlap within the same time interval. Such corrections are expected to be subleading with respect to the double logarithmic ones.}. The corresponding correction $\SIII$  can then be written as follows
\beq\label{DS3}
\Delta \SIII_{a_0b_0c_0} (t_1,t_0)&=& \frac{\alpha_s}{N_c^2-1}\int \frac{d\Romega}{\Romega^3} \int _{t_0}^{t_1}\rmd t_3 \int _{t_0}^{t_3}\rmd t_2  \nn  &\times&  \SIII_{a_3b_3c_3} (t_1,t_3)\, F_{a_3b_3c_3,a_2b_2c_2} (t_3,t_2)\,\SIII_{a_2b_2c_2,a_0b_0c_0} (t_2,t_0)\, ,\nn
\eeq
where we have factorized the contributions of the various time intervals, $[t_0,t_2]$, $[t_2,t_3]$ and $[t_3,t_1]$. These contributions are written as color tensors. 
The tensor $F$ is associated with the intermediate 4-point function (see Fig.~\ref{rad-3p-A3}), and the other two tensors are attached to 3-point functions, with 
\beq\label{3-pt-f}
\SIII_{a_3b_3c_3}\equiv f^{a_1b_1c_1}\SIII_{a_1b_1c_1,a_3b_3c_3}\,,\qquad  \SIII_{a_3b_3c_3} (t_1,t_3)\equiv f_{a_3b_3c_3}  \SIII (t_1,t_3)\,,
\eeq
where $\SIII (t_1,t_3)$ is the scalar 3-point function defined in Eq.~(\ref{3-pt-def}). 
Note that the choice of the color indices  $a_i,b_i,c_i$ is correlated to that of the coordinates $\y_i,\x_i,\bar\x_i$, with $i=0,1,2,3$ labeling the various times.  
We are leaving  open the color indices at the initial time $t_0$, and we shall  show that $\Delta\SIII_{a_0b_0c_0}$ is proportional to  $f^{a_0b_0c_0}$. This will guarantee  that the color structure exhibited in Eq.~(\ref{DS3}) will iterate itself when multiple radiative corrections are included before time $t_0$, similarly to what happens to the 3-point function in the absence of radiative corrections (cf. Eq.~(\ref{integralequation4})).

We shall denote with parenthesis the contributions to $F$ of the various diagrams, writing
\beq
 (X_3|F|X_2)\equiv \delta(\y_3-\y_2)\delta(\x_3-\x_2)\delta(\bar \x_3-\bar \x_2)\sum_{i=1}^3\left[(A_i)+(B_i)+(C_i)\right]\,.
\eeq
  For instance, the contribution to $F$ of  diagram $A_3$ in Fig.~\ref{rad-3p}, which we denote by $(A_3)$, reads
\beq\label{DS3-A3}
(A_3 )_{a_3b_3c_3,a_2b_2c_2} \equiv \langle \left(U^{\dag}_{\bar \x_3}T^{\dag e_3}\right)^{c_2c_3}U^{b_3b_2}_{\x_3} \left(U_{\y_2}T^{e_2}\right)^{a_3a_2}\bdel_{\z_2}\cdot\bdel_{\z_3}(\z_3|\Gc^{e_3 e_2}|\z_2)\Big|_{\z_2=\y_2,\z_3=\bar\x_3}\rangle\,,\nn
\eeq
where we have used the fact that during the radiation time, $\tau'=t_3-t_2$, the transverse coordinates of the emitting gluons are frozen in order to replace  their propagators  by trivial Wilson lines, e.g., $(\y_3|{\cal G}(t_3,t_2)|\y_2)\approx \delta(\y_3-\y_2)\,U_{\y_2}(t_3,t_2)$ for the gluon with frequency $\omega$, and similarly for the two other emitters with frequency $E$. The two derivatives acting on the coordinates of the radiated gluon at the end points of its propagation, namely $\bdel_{\z_2}$ and $\bdel_{\z_3}$, originate from the eikonal vertices (other factors, such as $g$ and $1/z$ have been factored out).  Often in what follows, we shall set $ \y_2=\y_3\equiv \y$, $\x_2=\x_3\equiv \x $ and $\bar\x_2=\bar\x_3\equiv \bar\x $ in the arguments of the Wilson lines, and we shall not indicate the dependence on the time variables in order to simplify the formulae.  

We proceed now to the systematic analysis of the color structure of the 9 diagrams that contribute to Eq.~(\ref{DS3}). These  differ only in the transverse coordinates of the emission and absorption of the softer gluon. We do the analysis starting from the right parts of the diagrams, i.e., from the latest times, and then moving to the left, i.e., to the early times (see  Fig.~\ref{rad-3p-A3}).
The 3-point function  without radiative corrections, $ \SIII_{a_3b_3c_3} (t_1,t_3)$, that ends at time $t_1$ and is common to all nine diagrams, obeys the second equality in Eq.~(\ref{3-pt-f}). 
By  contracting  $f_{a_3b_3c_3}$ with the 4-point  tensor $F$ in the region  $[t_2,t_3]$, we get
\beq\label{4-point}
(A_3)_{a_2b_2c_2}&\equiv&f^{a_3b_3c_3}\,(A_3)_{a_3b_3c_3,a_2b_2c_2}\nn
&=& f_{a_3b_3c_3} \langle \left(U^{\dag}_{\bar \x}T^{ e_3}\right)^{c_2c_3}U^{b_3b_2}_{\x} \left(U_{\y}T^{e_2}\right)^{a_3a_2}\bdel_{\z_2}\cdot\bdel_{\z_3}(\z_3|\Gc^{e_3 e_2}|\z_2)\Big|_{\z_2=\y_2,\z_3=\bar\x_3}\rangle,\nn
\eeq
where we have chosen the diagram $A_3$ as an example. 
The complete result for $F$  is obtained by summing over all possible hookings of the radiated gluon with the requirement that the gluon propagator is ${\cal G}(t_2,t_3)$ if it initiates at $\x_2$ or $\y_2$ and ${\cal G}^\dag(t_3,t_2)$  if it initiates at $\bar\x_2$, as is the case for the contributions $C$. \\

At this point, we invoke the fact that $\omega\ll E$, as already mentioned.  According to Eq.~(\ref{freecalGx}), the soft gluon $\omega$ therefore explores transverse distances, that are large compared to the sizes of the emitting dipole.  Given this, one can neglect, in the 4-point function that appears in the interval $[t_2,t_3]$, the dipole size of the hard gluon, $\v \equiv \x-\bar\x$, with respect to that of the  soft gluon $\u\equiv \y-\x$. In fact, in the calculation of the spectrum, there is no approximation involved here: indeed, the integration over $\l$ (conjugate of $\v$ in Eq.~(\ref{3-point-tilde}))  in Eq.~(\ref{kernel5}) yields a factor $\delta(\v)$ which enforces the size of this hard gluon dipole to vanish.

\begin{figure}[htbp]
\centering
 \includegraphics[width=5.5cm]{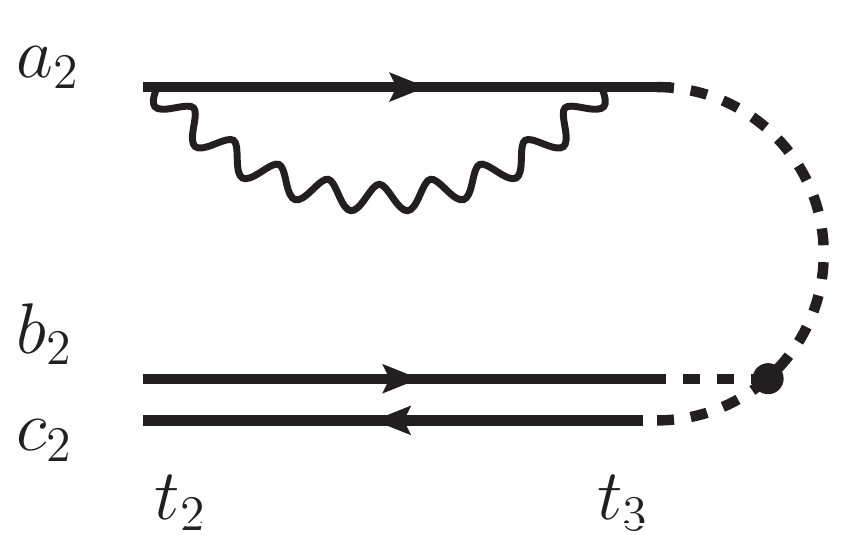}\qquad\qquad  \includegraphics[width=5.5cm]{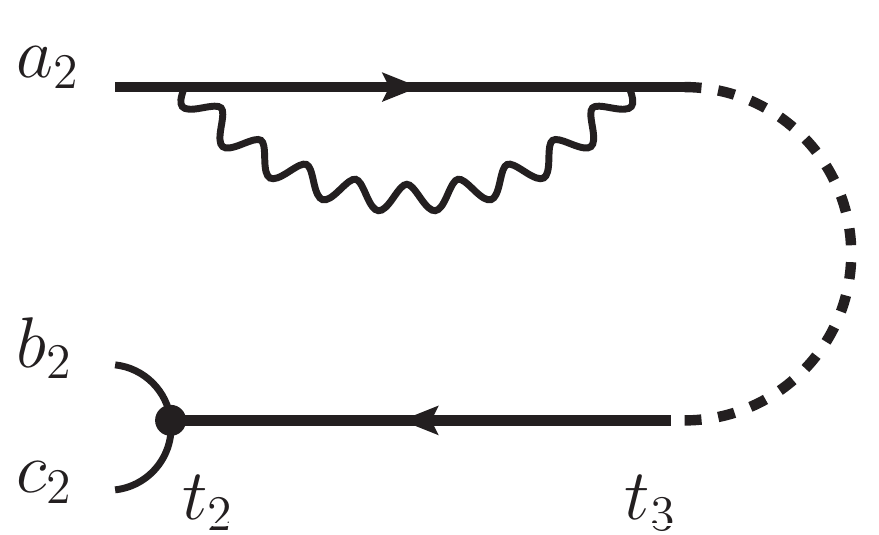} \\
  \includegraphics[width=9cm]{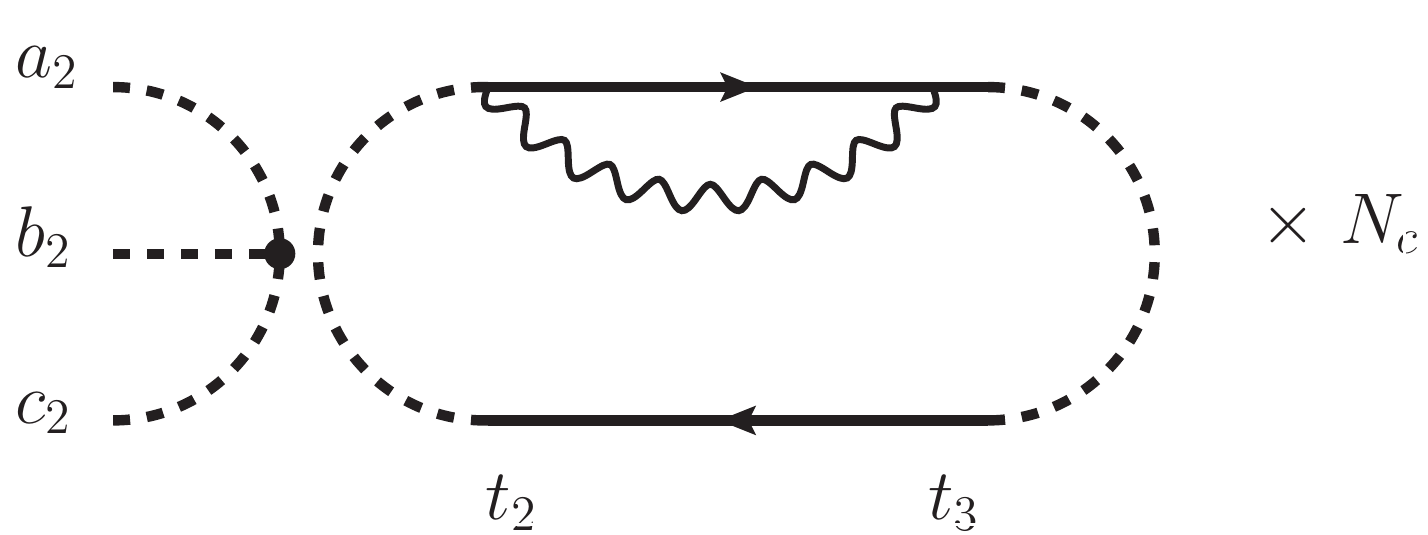}

\caption{$(A_1)$ contribution to the radiative correction to the 3-point function. The three diagrams represent various steps in the color algebra. The lower diagram represents the final factorized result in Eq.~(\ref{Fierz1}). The thick lines represent Wilson lines and the wavy line  the radiated gluon. The dashed lines in the r.h.s of the first diagram indicates that the Wilson lines are coupled to an $f$ symbol represented by a black dot. In the second diagram, the generalized Fierz identity (\ref{Fierz2}) has been used to move the dotted vertex to the time $t_2$ reducing the two Wilson lines at equal transverse position $\x$ to a single Wilson line which, together with the upper Wilson line (at position $\y$), is coupled to a singlet. The last diagram represents the last step in the color reduction: the two Wilon lines at $t_2$ are in a color singlet, and the $f^{a_2b_2c_2}$ symbol emerges as the remaining color structure. }
\label{fig9virt}
\end{figure}

\begin{figure}[htbp]
\centering 
 \includegraphics[width=5.5cm]{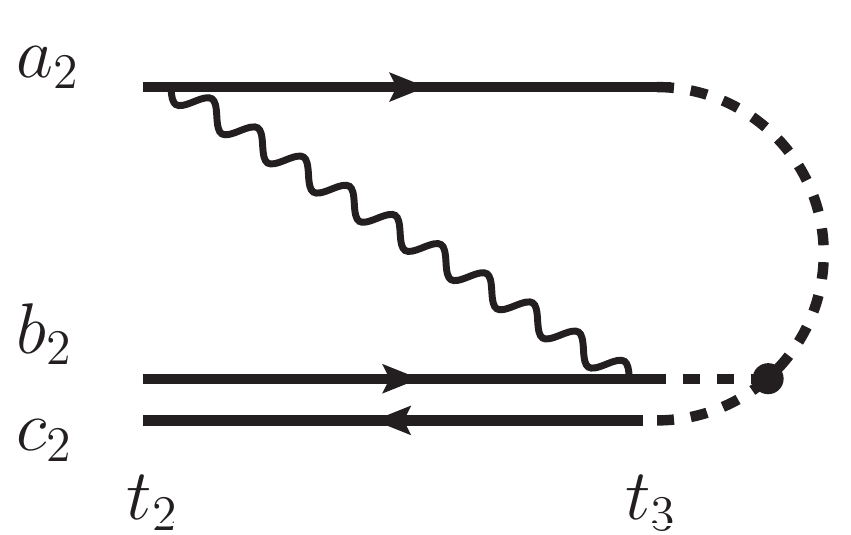} \qquad\qquad \includegraphics[width=5.5cm]{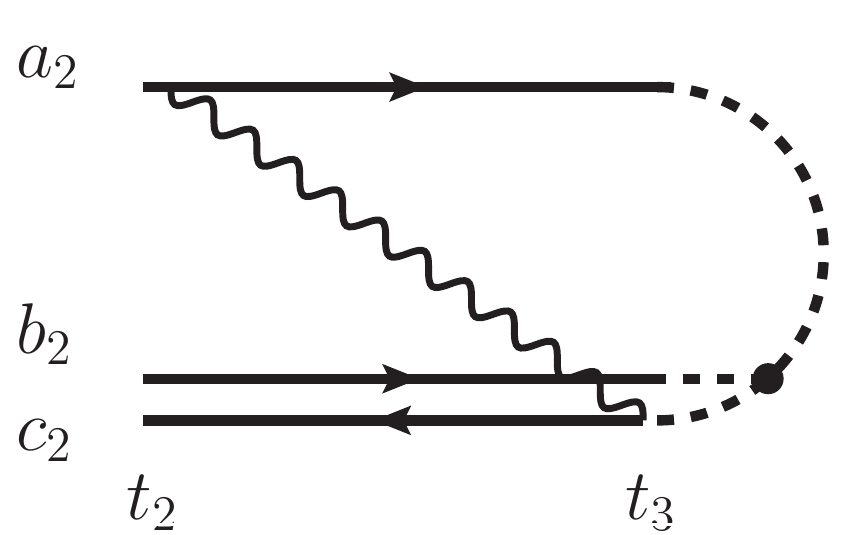}\vspace{0.3cm}
  \includegraphics[width=9cm]{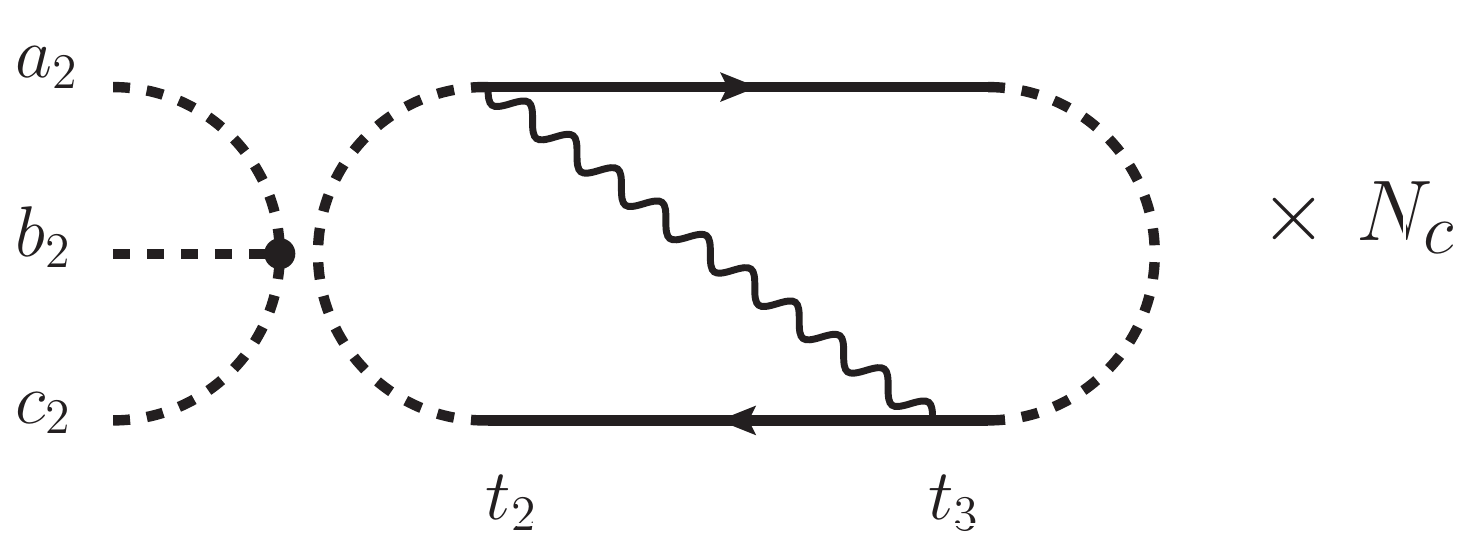} 

\caption{Upper panel: $(A_2)$ (left) and $(A_3)$ (right) contributions.  Lower panel: the result for $(A_2)+(A_3)$. Combining the upper two diagrams using the Fierz identity (\ref{Fierz}) allows us to move the dotted vertex to the left of the soft gluon endpoint (wavy line) at time $t_3$. Then the extended form of the Fierz identity, Eq.~(\ref{Fierz2}),  allows us to move the dotted vertex  to the time $t_2$. As a result the two lower Wilson lines collapse into a single one, and we end up with two Wilson lines at transverse coordinates $\y$ and $\x$,  coupled to a singlet at both $t_3$ and $t_2$. The final result is a reduced 3-point function multiplied by $N_c f^{a_2b_2c_2}$.}
\label{fig9}
\end{figure}

Exploiting this property ($\bar\x=\x$), we can write the total contribution  $(A)=(A_1)+(A_2)+(A_3)$ as follows
\beq\label{A-cont}
(A)_{a_2b_2c_2}&=& -i\left\langle\left[-\left(U^{\dag}_\x T^{a_3}U_\x\right)_{c_2b_2} \left(T^{e_3}U_\y T^{e_2}\right)_{a_3a_2}\right]\bdel_{\z_2}\cdot\bdel_{\z_3}(\z_3|\Gc^{ e_3 e_2}|\z_2)\Big|_{\z_2=\z_3=\y}\right.\,\nn
&&\;\, \;-i\left[-\left(U^{\dag}_{\x}T^{a_3}T^{ e_3}U_\x \right)_{c_2b_2} \left(U_\y T^{e_2}\right)_{a_3a_2}\right.\nn
&&\qquad \left.\left.+\left(U^{\dag}_\x T^{e_3}T^{a_3}U_\x\right)_{c_2b_2} \left(U_\y T^{e_2}\right)_{a_3a_2}\right]\bdel_{\z_2}\cdot\bdel_{\z_3}(\z_3|\Gc^{e_3 e_2}|\z_2)\Big|_{\z_2=\y,\z_3=\x} \right\rangle,\nn
\eeq
where used has been made of the identity $(T^a)_{cb}=if^{abc}$. Note that the first two terms, $A_1$  and $A_2$, are virtual corrections where the gluon is emitted and reabsorbed in the amplitude. This is the reason why they come with a relative minus sign as compared to $A_3$.
These three contributions are illustrated in the first diagram of Fig. \ref{fig9virt} for the term $(A_1)$, and the first two diagrams in Fig. \ref{fig9}, for the last two terms $(A_2)$ and $(A_3)$.

The last two terms in Eq.~(\ref{A-cont}), corresponding to the contributions $(A_2)$ and $(A_3)$, can be given a form similar to that of the first term $(A_1)$, by  using the Fierz identity 
\beq\label{Fierz}
T^{a}T^{e}-T^{e}T^{a}=[T^{a},T^{e}]=if^{a e d} T^d=-(T^{ e})_{da} T^d.
\eeq
By combining the three contributions, one then gets 
\beq\label{Fierz1}
&& -i\left[-\left(U^{\dag}_\x T^{a_3}U_\x \right)_{c_2b_2} \left(T^{e_3}U_\y T^{e_2}\right)_{a_3a_2}\right]\bdel_{\z_2}\cdot\bdel_{\z_3}(\z_3|\Gc^{e_3 e_2}|\z_2)|_{\z_2=\z_3=\y}\,,\nn
&&-i\left[+\left(U^{\dag}_\x T^{a_3}U_\x \right)_{c_2b_2} \left(T^{e_3}U_\y T^{e_2}\right)_{a_3a_2}\right]\bdel_{\z_2}\cdot\bdel_{\z_3}(\z_3|\Gc^{e_3 e_2}|\z_2)|_{\z_2=\y,\z_3=\x}\,,\nn
&&=-i\left(U^{\dag}_\x T^{a_3}U_\x \right)_{c_2b_2} \left(T^{e_3}U_\y T^{e_2}\right)_{a_3a_2}\nn
&&\qquad\times\bdel_{\z_2}\cdot\bdel_{\z_3}\left[(\z_3|\Gc^{e_3 e_2}|\z_2)|_{\z_2=\y,\z_3=\x}-(\z_3|\Gc^{e_3 e_2}|\z_2)|_{\z_2=\z_3=\y}\right]\,.
\eeq
We now make use of the following extended form of the Fierz identity \beq\label{Fierz2}
\left(U^{\dag}_\x T^{a_3} U_\x \right)_{c_2b_2}=(T^d)_{c_2b_2} (U^\dag_\x)_{da_3}\,,
\eeq
and we get 
\beq\label{S4}
&&(A)_{a_2b_2c_2}=-i\left(T^{d}\right)_{c_2b_2} \langle\left(U^{\dag}_\x T^{e_3}U_\y T^{e_2}\right)_{da_2}\nn
&&\qquad\qquad\qquad\times\bdel_{\z_2}\cdot\bdel_{\z_3}\left[(\z_3|\Gc^{e_3 e_2}|\z_2)|_{\z_2=\y,\z_3=\x}-(\z_3|\Gc^{e_3 e_2}|\z_2)|_{\z_2=\z_3=\y}\right] \rangle\,.
\eeq
Thus, the 4-point function reduces to a 3-point function which has a trivial color structure, that is, it is proportional to $f^{a_2b_2c_2}$. Hence, 
\beq\label{S4-2}
&&(A)_{a_2b_2c_2}= N_c \,f^{a_2b_2c_2}  \nn
&&\qquad \bdel_{\z_2}\cdot\bdel_{\z_3}\left[\tiSIII(\z_2-\y,\z_3-\y,\u;\tau')\Big|_{\z_2=\y,\z_3=\x} -\tiSIII(\z_2- \y,\z_3-\y,\u;\tau')\Big|_{\z_2=\z_3=\y} \right]\,,\nn
\eeq
where we have recognized the reduced 3-point function of Eq.~(\ref{3-point-tilde2}), 
\beq
\tiSIII(\z_2-\y,\z_3-\y,\u;\tau')=\frac{1}{N_c(N_c^2-1)}\langle \text{Tr}\left(U^{\dag}_\x T^{e_3}U_\y T^{e_2}\right)\,(\z_3|\Gc^{e_3 e_2}|\z_2)\rangle,
\eeq
whith $\u=\y-\x$ and $\tau'=t_3-t_2$. 
Using the definition in Eq.~(\ref{K-x}), we get for the real part of  $(A)_{a_2b_2c_2}$ 
\beq
{\rm Re}(A)_{a_2b_2c_2}= N_c \,f^{a_2b_2c_2}\frac{1}{2} K(\u, \tau')\,,
\eeq 
which contains the double logarithmic enhancement encoutered in Eq.~(\ref{dsig-0}). 

The various manipulations that we have performed can be understood graphically. This is illustrated in Fig. \ref{fig9virt} and \ref{fig9}, and we refer to the captions of these figures for further details on the calculation.

\begin{figure}[htbp]
\begin{center}
 \includegraphics[width=5.5cm]{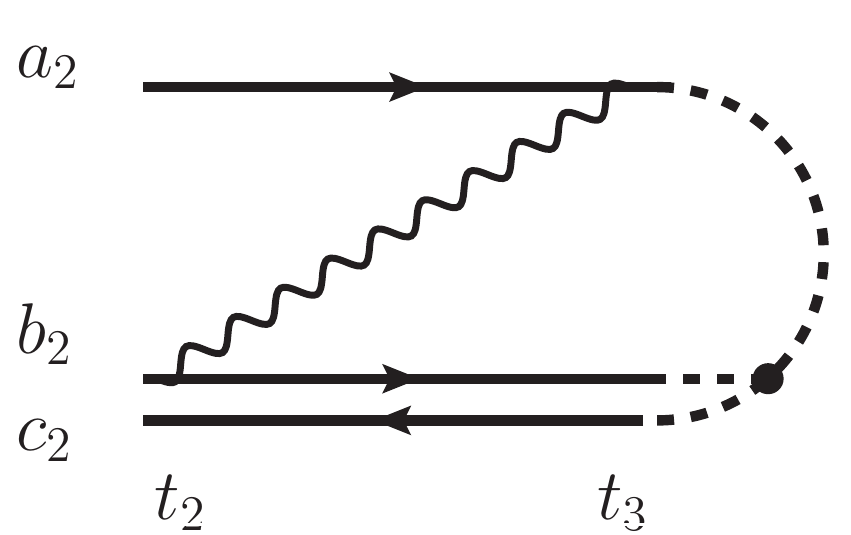}  \qquad\qquad\includegraphics[width=5.5cm]{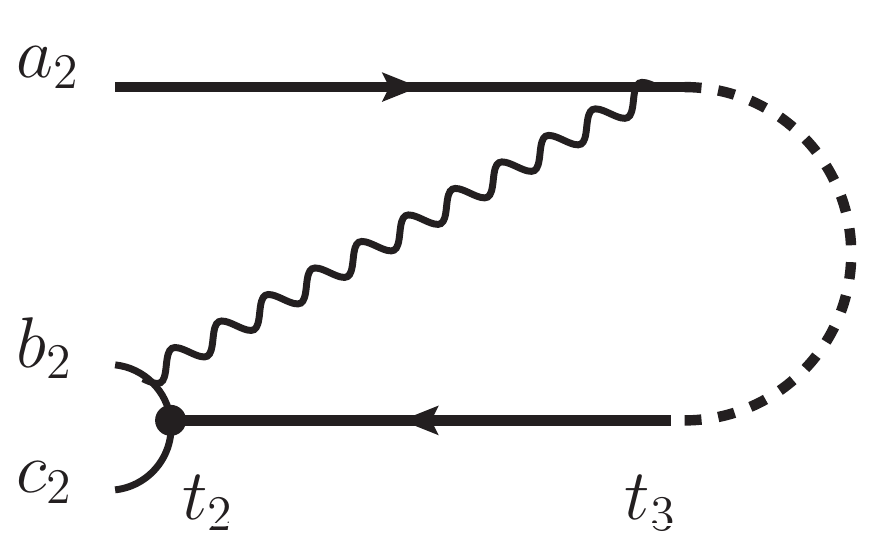}\vspace{5mm} \includegraphics[width=9cm]{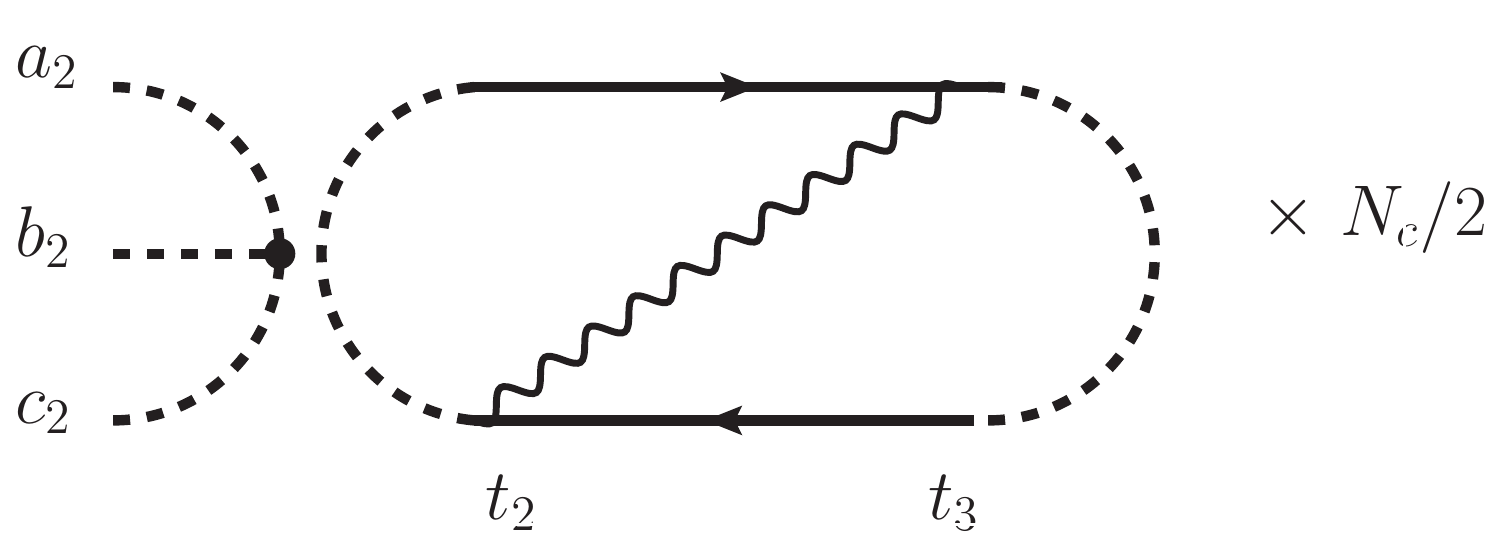} 

\caption{ The three steps in the color algebra analogous to those in  Fig. \ref{fig9}, here for the $B_1$ contribution. }
\label{fig10}
\end{center}
\end{figure}

\begin{figure}[htbp]
\begin{center}
 \includegraphics[width=5.5cm]{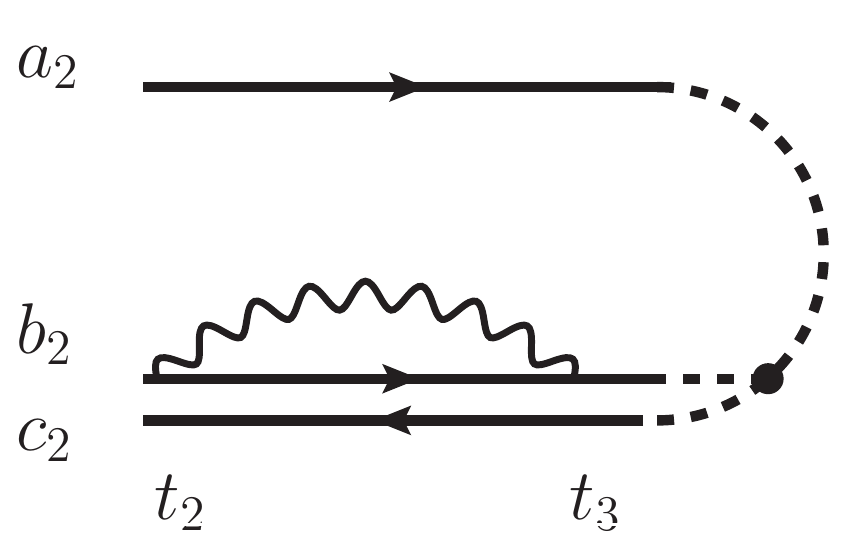}  \qquad\qquad\includegraphics[width=5.5cm]{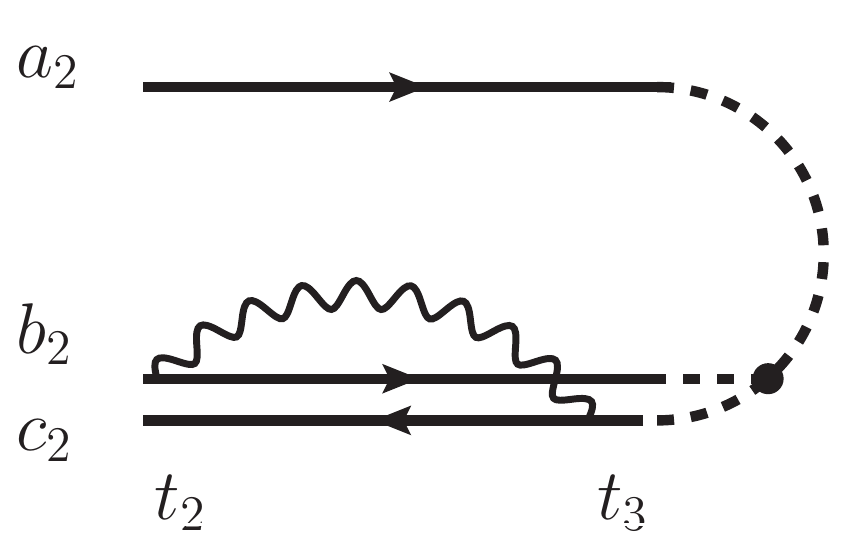} \\ \includegraphics[width=9cm]{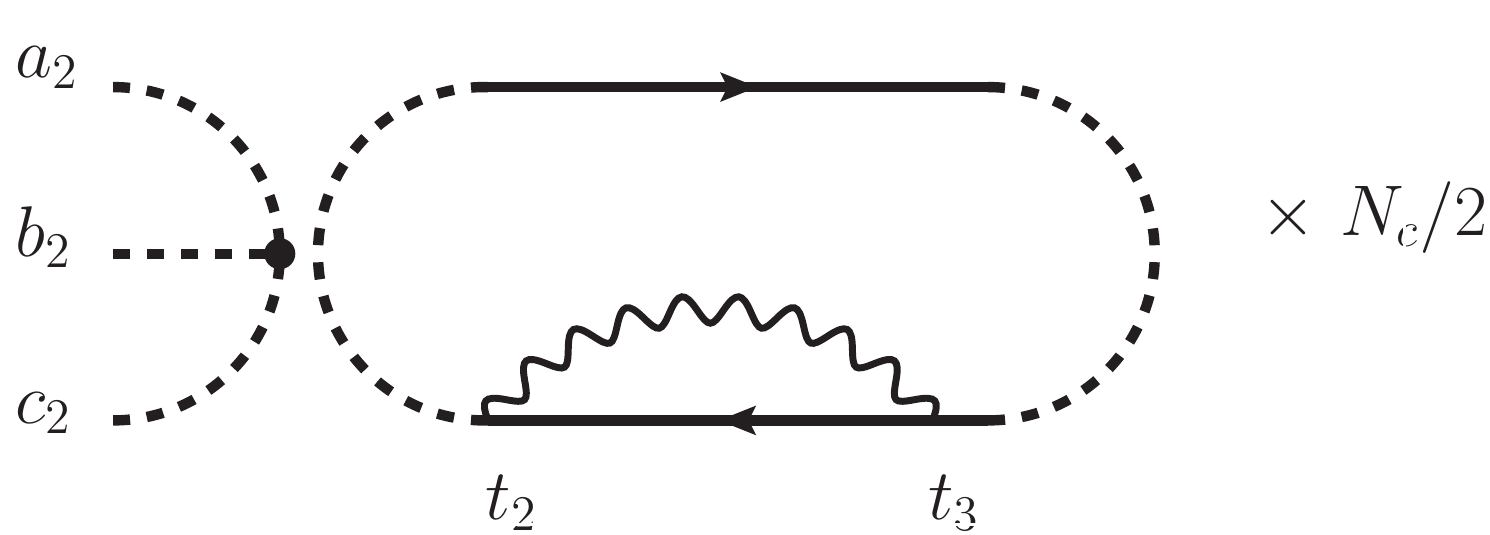} 

\caption{ The three steps in the color algebra analogous to those in  Fig. \ref{fig9}, here for the $B_2+B_3$ contribution.  }
\label{fig11}
\end{center}
\end{figure}

We still have to account for the emission off the lower lines, corresponding to the diagrams  depicted partly in Figs. \ref{fig10} and \ref{fig11}. We shall be very brief here, since the manipulations to be done involve the same techniques as used for the diagrams $A$, namely the use of Fierz identities to rearrange terms. Illustrations of the procedure are given in Figs.~\ref{fig10} and \ref{fig11}.   Consider diagrams $B$. Using the Fierz identity for Wilson-lines we reduce diagram $B_1$ to a 3 point function with color factor $N_c/2$ times $f^{a_2b_2c_2}$. Diagrams $B_2$ and $B_3$ also combine  into a 3-point function with color factor $N_c/2$. Diagrams $C$ can be computed similarly and yield the complex conjugate of the 3-point fonction since it involves a ${\cal G}^\dag$. It follows that the sum of diagrams $B$ and $C$ is real, and given by 
\beq\label{S4-3}
&&(B+C)_{a_2b_2c_2}=  \frac{1}{2(N_c^2-1)} \,f^{a_2b_2c_2}  \bdel_{\z_2}\cdot\bdel_{\z_3}\nn
&& \left\{ \langle{\rm Tr}\left(U^{\dag}_\x T^{e_3}U_\y T^{e_2}\right)\left[(\z_3|\Gc^{e_3 e_2}|\z_2)|_{\z_2=\x,\z_3=\y}-(\z_3|\Gc^{e_3 e_2}|\z_2)|_{\z_2=\z_3=\x}\right] \rangle\right.\nn
&&\left.+ \langle{\rm Tr}\left(U^{\dag}_\x T^{e_3}U_\y T^{e_2}\right)\left[(\z_3|\Gc^{\dag e_3 e_2}|\z_2)|_{\z_2=\x,\z_3=\y}-(\z_3|\Gc^{\dag e_3 e_2}|\z_2)|_{\z_2=\z_3=\x}\right] \rangle\right\}\,.\nn
\eeq
Using the fact that ${\rm Tr}\left(U^{\dag}_\x T^{e_3}U_\y T^{e_2}\right)={\rm Tr}\left(U_\x T^{e_3}U^{\dag}_\y T^{e_2}\right)$ is real, it follows that the sum of diagrams $B$ and $C$ is also real, and given by 
\beq\label{S4-4}
&&(B+C)_{a_2b_2c_2}= N_c \,f^{a_2b_2c_2} \nn
&&\text{Re}\, \bdel_{\z_2}\cdot\bdel_{\z_3}\left[\tiSIII(\z_2-\x,\z_3-\x,-\u,\tau')\Big|_{\z_2=\x,\z_3=\y} -\tiSIII(\z_2-\x,\z_3-\x,-\u,\tau')\Big|_{\z_2=\z_3=\x} \right], \nn\eeq
which, by using  the symmetry of $K(\u,\tau')$  in the exchange $\u\to -\u$, we can write as 
\beq\label{S4-5}
&&(B+C)_{a_2b_2c_2}=\frac{1}{2} N_c \,f^{a_2b_2c_2} K(\u,\tau')\,.
\eeq

The correction to the 3-point function (in the limit $\v\to\0$ as required by the integration over $\l$ in Eq.~(\ref{kernel5})) can now be written by analogy with Eq.~(\ref{integralequation4}) as 
\beq\label{DS3-2}
\Delta \SIII  (t_1,t_0)\simeq - \int _{t_0}^{t_1}\rmd t_3 \int _{t_0}^{t_3}\rmd t_2   \SIII (t_1,t_2) \, \Delta\Sigma^{(3)}(t_3,t_2)\,\SIII (t_2,t_0)\,,
\eeq
where 
\beq
(X_3|\Delta\Sigma^{(3)}(t_3,t_2)|X_2)=\delta(X_3-X_2)\frac{\alpha_s N_c}{2}\int \frac{d\omega'}{\omega'^3}K(\u,\tau')\,.
\eeq
At this point we proceed as for the 2-point function, and treat the correction $\Delta\Sigma^{(3)}(t_3,t_2)$ as a local correction. By comparing with Eq.~(\ref{Sigma3a}) one then gets
\beq\label{Deltasigma3}
\frac{N_c n}{4}[\Delta\sigma(\y-\bar\x)+\Delta\sigma(\y-\x)]=\frac{N_c n}{2}\Delta\sigma(\u)=\frac{\alpha_s N_c}{2}\int^{\tau_\text{max} } \rmd \tau'\int \frac{\rmd\omega'}{\omega'^3}K(\u,\tau')\,,
\eeq
in agreement with Eq.~(\ref{dsig-1}). We have used $\Delta\sigma (\y-\x)=\Delta\sigma(\y-\bar\x)=\Delta\sigma(\u)$, and $\Delta\sigma(\x-\bar\x)=0$.\\

We have thus shown that to double logarithmic accuracy, the radiative corrections to the reduced 3-point function are accounted for by correcting the dipole cross-section and thus  the jet quenching parameter.  This result applies to the particular 3-point function involved in the calculation of the gluon spectrum (\ref{kernel5}). In order to evaluate the corrected spectrum, we need to perform the integration over  $\tau$ in  Eq.~(\ref{kernel5}), replacing $\hat q$ by $\hat q +\Delta\hat q (\tau_\text{max})$. However, since the correction to $\hat q$ is computed to double logarithmic accuracy one can simply replace the variable $\tau_\text{max}$ in Eq.~(\ref{Deltasigma3}) (see also Eq.~(\ref{qhat1})) by its typical value in the radiation process, i.e., $\tau_\text{max}\simeq \tau_\text{br}\equiv \sqrt{\omega / \hat q}$. The integral over $\tau$ in the BDMPS spectrum (\ref{kernel5}) can then be perfumed as for the case with no radiative correction. Doing so, we obtain
\beq\label{BDMPS-rad}
\frac{\rmd N}{\rmd\omega \rmd t}\equiv \frac{\alpha_s N_c}{\pi } \sqrt{\frac{\hat q+\Delta \hat q}{\omega}}\,,
\eeq
where for a constant $\hat q$ one gets, from Eq.~(\ref{qhat1}) letting $\p^2\simeq k^2_\text{br}(\omega)\equiv \sqrt{\omega\hat q} $,
\beq
\hat q+\Delta \hat q \approx \hat q\left[1+\frac{\alpha_sN_c}{2\pi}\ln^2\sqrt{\frac{\omega}{\hat q \tau^2_0}}\right]\,. 
\eeq
We shall return to this formula in the discussion section below. 
\subsection{General 3-point function in the large $N_c$ limit}

In the previous subsection, we  have shown that the radiative corrections to the BDMPS spectrum can absorbed in a renormalization of the quenching parameter. This result was obtained for the special  where the radiated gluon is soft,  $(1-z) \to 0 $,  and the size of the  dipole made by  the energetic gluons in the amplitude and the complex conjugate amplitude vanishes, i.e., $\v=\0$. In this special case, the color algebra, which constitutes the main technical difficulty of the calculation, simplifies considerably. There is another situation where the color algebra simplifies greatly, this is the limit of large $N_c$. In this limit, as we now show, it is possible to generalize the previous  result concerning the renormalization of $\hat q$ to the general 3-point function that describes the medium induced splitting, and which controls the in-medium QCD cascade \cite{Blaizot:2013vha} .

 \begin{figure}[htbp]
\begin{center}
 \includegraphics[width=12cm]{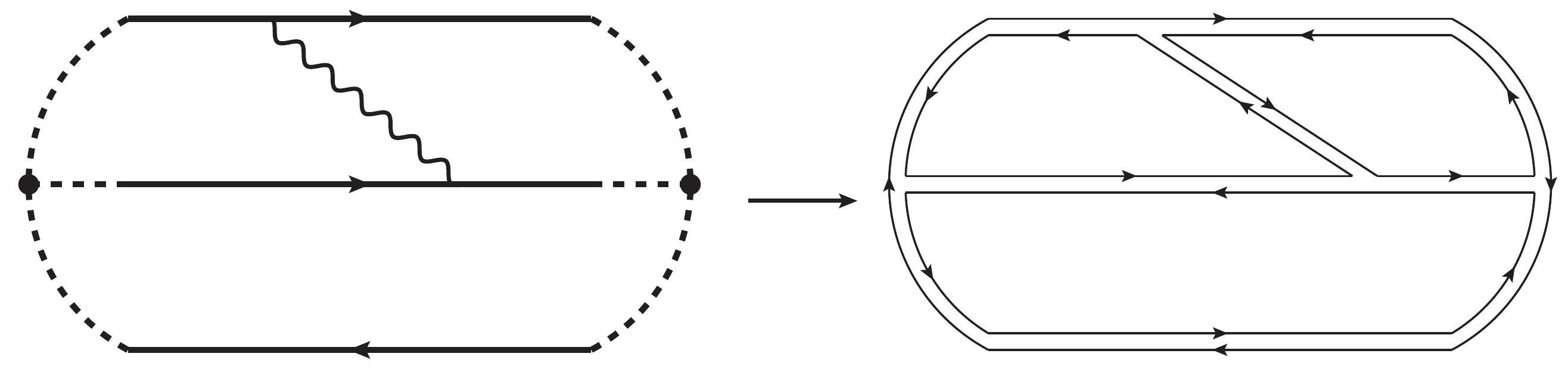} 
\caption{ Representation of the 3-point function with one radiative correction. In the right panel, the double line representation that facilitates the analysis of the  large $N_c$ limit. }
\label{fig12}
\end{center}
\end{figure}

  The representation of the 3-point function in the large $N_c$ limit is given in Fig. \ref{fig12}, where in the right hand side, we have used the standard double line notation to represent the gluons. The diagrams in  Fig. \ref{fig12} represent a single radiative correction, and the interaction with the medium is ignored at this point (it will be discussed shortly). As made clear by the double line notation of the diagram on the right, the additional gluon represents a correction of order $g^2 N_c$, where the factor $N_c$ comes from the additional loop.

\begin{figure}[htbp]
\begin{center}
 \includegraphics[width=12cm]{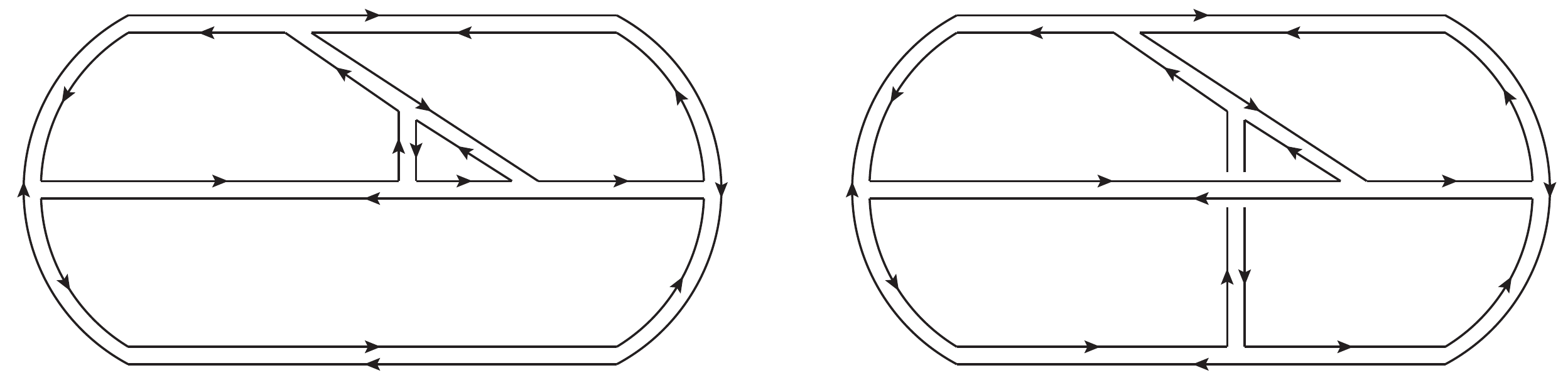} 
\caption{ The dominant contribution in the large $N_c$ limit is given by the graphs where the three dipoles remains disconnected. Left panel: planar graph of  order $(g^2 N_c)(nN_c)$, involving the radiated gluon and one instantaneous medium interaction. Right panel: a non planar graph, which is suppressed by a factor $1/N_c^2$ with respect the planar diagram. }
\label{fig14}
\end{center}
\end{figure}

  Consider now  the effect of a single medium interaction (which is the dominant contribution to the double logarithmic singularity). The corresponding diagrams are displayed in Fig.~\ref{fig14}. The medium interaction is represented by vertical line to reflect the instantaneity  of the medium interaction (see Fig.~\ref{fig:fig2}). There are two distincts contributions. On the left hand side of Fig.~\ref{fig14}, the medium interaction takes place in the same dipole as that involved in the radiative correction, the  dipole $\y\x$.  This diagram yields a factor $g^2N_c (nN_c)$, where the factor $nN_c$ comes from the instantaneous interaction (see Eq.~(\ref{sigmaSigma2})). This factor $nN_c$ is absorbed in $\hat q$, while the factor $g^2N_c$ is the factor that comes multiplying the double logarithmic correction. The diagram in the right hand side of Fig.~\ref{fig14} on the other hand, has a medium interaction connecting two distinct dipoles, the dipole $\y\x$ and the dipole $\x\bar\x$. It is a non planar diagram, whose contribution is suppressed by a factor $1/N_c^2$ with respect to that of the left diagram. 

This argument is easily extended to all the diagrams that are relevant for the calculation of the double logarithmic correction. One finds that the leading contributions at large $N_c$ are those which involve each single dipole individually. In other words, in the large $N_c$ limit, each individual dipole receives an independent correction, which can then be reabsorbed as a correction to the corresponding  dipole-cross section in Eq.~(\ref{3-point-tilde-xperp}):
\beq
&&\sigma(\y-\x)\,\to\, \sigma(\y-\x)+\Delta\sigma(\y-\x)\,,\nn
&&\sigma(\x-\bar\x)\,\to\, \sigma(\x-\bar\x)+\Delta\sigma(\x-\bar\x)\,,\nn
&&\sigma(\y-\bar\x)\,\to\, \sigma(\y-\bar\x)+\Delta\sigma(\y-\bar\x)\,.\nn
\eeq

We have then generalized our result to the general 3-point function and therefore, to the branching kernel and the full medium-induced jet evolution in the large $N_c$ limit. 
Note that in the case examined in the previous subsection, $\sigma(\x-\bar\x)=0=\Delta\sigma(\x-\bar\x)$, so that we recover the result of the previous section. It is remarkable that the general result is proportional to  $N_c$ and hence coincides with that of the large $N_c$ limit.

\section{Discussion}\label{discussion}
We end this paper by discussing several implications of the scale dependence of the jet-quenching parameter coming from the radiative corrections. 

\subsection{Radiative correction to the mean energy loss}
Let us first consider the typical transverse momentum squared. This  is computed using the broadening probability at $t=L$,
\beq\label{k2P}
\langle k_\perp^2\rangle_\text{typ}\equiv \int_\k \k^2 \, {\cal P} (\k,L)\,.
\eeq
Since the renormalized jet quenching parameter is computed to logarithmic accuracy, this typical transverse momentum squared is sufficient to set the typical scale in the logarithms,  i.e., $\hat q (L,\k) \sim \hat q (L,\langle k_\perp^2\rangle_\text{typ})$. Under this condition,  the solution of the diffusion equation is Gaussian. One recovers from Eq.~(\ref{P-mom-FP}) the resulted obtained already  in \cite{Liou:2013qya}, 
\beq\label{meank2}
\langle k_\perp^2\rangle_\text{typ}\simeq  \hat q L  \left (1+\frac{\bar \alpha}{2}  \ln^2\frac{L}{\tau_0}\right)\,.
\eeq
where we have replaced $\hat q$ by $\hat q+\Delta \hat q(\tau_\text{max}, 1/\langle \x^2\rangle)$ and set $\tau_\text{max}\equiv L$ and $1/\langle \x^2\rangle\sim \langle k_\perp^2\rangle_\text{typ}\sim \hat q L$ (to this level of accuracy it is sufficient to put the leading order $\hat q$ in the argument of the logs) in Eq.~(\ref{deltaqhat}). Here $\bar\alpha\equiv \alpha_s N_c/\pi$.

Similarly, by replacing $\hat q $ by its corrected value $\hat q+\Delta\hat q(\tau_\text{br}(\omega),k^2_\text{br}(\omega))$  in Eq.~(\ref{BDMPS}), on can compute the correction to the mean energy loss. Since the integral is dominated by gluons with formation times of the order of the length of the medium, $\omega\sim \omega_c$, we shall set, as for the transverse momentum squared, $\tau_\text{br}\sim L$ and  $k^2_\text{br}\sim \hat q L$. Therefore, we readily obtain, to logarithmic accuracy, 
\beq
\langle \omega \rangle \sim  \hat q L^2 \left (1+\frac{\bar \alpha}{2}  \ln^2\frac{L}{\tau_0}\right)\,,
\eeq
where the corrected jet-quenching parameter is identical to that in Eq.~(\ref{meank2}).

\subsection{Renormalization of the jet-quenching parameter}
For large media, as soon as  $\bar\alpha\ln^2(L/\tau_0)\sim1$ one has to resum the double logarithmic power corrections. Unlike the previous resummation of independent multiple radiative corrections, this now involves radiative corrections that are correlated to each other. To understand how this resummation proceeds, we denote the standard leading order definition of the jet-quenching parameter $\hat q\equiv\hat q_0$ and we note that the first correction to the jet-quenching parameter, $\hat q_1(\tau,\k^2)\equiv \Delta \hat q(\tau,\k^2)$ is proportional to the 3-point function, $S^{(3)}[\hat q_0]$ which is itself  a function of the leading order $\hat q_0$.  As we have shown, the 3-point function gets renormalized as $S^{(3)}[\hat q_0]\to S^{(3)}[\hat q_0+\hat q_1] $. This allows us to compute the second correction from Eq.~(\ref{qhat1}),
\beq
\hat q_2(\tau,\k^2)= \bar\alpha \, \int^{\tau }_{\tau_0} \frac{\rmd\tau'}{\tau'}\int^{1/\x^2}_{\hat q_0\tau'}\frac{\rmd\q^2}{\q^2}\,\hat q_1(\tau',\q^2)\,.
\eeq
This emerging self-similarity is a result of the separation of time scales involved in the computation of the leading logarithms.
The structure of the first double logarithmic corrections being set, the next corrections that yield double logarithms will follow the same systematics, with successive gluonic fluctuations ordered in formation time $\tau_0 \ll \tau _1\ll ...\ll\tau_n\equiv \tau_{\text{max}} $, or in transverse size $\x_0 \gg \x _1\gg ...\gg \x_n\equiv \x_{\text{max}} $, or in transverse momentum $m_D  \ll \q _1\ll ...\ll\q_n\equiv \k $. The difference with the standard Double-Logarithmic Approximation (DLA) is the limits of logarithmic phase-space set by the LPM effect since, i.e., multiple-scatterings since in the DLA only a single scattering contributes, which imposes that the formation time of a fluctuation to be smaller than the BDMPS formation time, or in terms of our transverse momentum variables,  $\q^2 \gg  \hat q_0\tau $. The following equation resums the  double logarithmic corrections to all orders
\beq\label{qhat-evol}
\frac{\del \hat q (\tau,\k^2)}{\del \ln \tau} = \, \int_{\hat q  \tau }^{\k^2}\frac{d\q^2}{\q^2}\, \bar \alpha(\q)\, \hat q (\tau,\q^2) \,.
\eeq
with some initial condition $ \hat q (\tau_0,\k)$. We have let the coupling running at the transverse scale $\q$.
Note the lower limit of the $\q$ integration which accounts for the boundary between single-scattering and multiple-scatterings. The important feature of this equation is it predicts the evolution of the jet-quenching parameter from an initial condition $\hat q_0$ (which can be computed e.g. on the lattice, or  to leading order in $\alpha_s$, which implies, $\hat q(\tau_0) \equiv\hat q_0  $ as given by the leading order result (\ref{qhat0}). The $\tau_0$ cut-off that was introduced  to cut the logarithmic divergence in the radiative corrections, can be seen as a factorization scale.  

The complete solution of the 2-dimensional evolution equation (\ref{qhat-evol}) is beyond the scope of this paper. Let us simply recall the solution derived in \cite{Liou:2013qya} for the pt-broadening in the case where $\hat q_0=\hat q(\tau_0)$ is constant and for a final $\tau=L$ and $\k^2=\hat q_0L$, merging the 2 independent variables at the end of the evolution. The solution reads,
\beq\label{qhat-L}
\hat q(L)=  \frac{1}{\sqrt{\bar\alpha}}\,I_1\left(2\sqrt{\bar\alpha}\ln\frac{L}{\tau_0}\right)\,\hat q(\tau_0)\,
\eeq
For large $L$, the quenching parameter scales like $\hat q(L)\sim L^\gamma $, with the anomalous 
\be\label{anomalous-d}
\gamma=2\sqrt{\bar\alpha}\,.
\ee
Interestingly, the resummation of large double logarithms modifies the scaling of the energy loss with $L$ the energy loss, $\langle \omega \rangle \sim L^{2+\gamma}$, and seems to fall between the standard small coupling result, $\langle \omega \rangle \sim L^2 $ and the strong coupling result obtained with the help of the AdS-CFT correspondance in ${\cal N}=4$ SYM theory , $\langle \omega \rangle \sim L^3 $ \cite{Hatta:2007cs}.

\section*{Acknowledgments}
We thank F. Dominguez, E. Iancu, A. H. Mueller and B. Wu  for numerous discussions on some of the issues discussed in this paper. E. Iancu has recovered some of the results of this paper form the point of view of evolution equations \cite{Iancu:2014kga}. Special thanks go to him for persuasive encouragements to speed up the writing of this paper. This research is supported by the European Research Council under the Advanced Investigator Grant ERC-AD-267258.

\appendix
\section{Calculational details}

In this appendix, we provide some additional details on the calculations that lead to the expressions in the main text, in particular the expression (\ref{deltaP}) for the modification, due to a radiative correction, of the branching probability of a gluon into two gluons. This is easily obtained through a straightforward, though tedious, calculation, starting from the results given in Ref.~\cite{Blaizot:2013vha}. We provide here some indication on how this calculation can be done, keeping at first the notation of Ref.~\cite{Blaizot:2013vha} to which we refer for further details. The calculation is based on that of the amplitude for a gluon (0) of momentum $\p_0$, color $c_0$ and polarization $\lambda_0$, present in the system at time $t_0$,  to evolve in the medium into a gluon $a$ with momentum $\k_a$, color $a$ and polarization $\lambda_a$ at time $t_L$, after emitting at a time $t_1$ ($t_0\leq t_1\leq t_L$) a soft gluon $b$ that is not observed, and is eventually integrated out. 
This amplitude can be written as
 \beq
{\cal M}^{abc_0}_{\lambda_a,\lambda_b,\lambda_0}(k_a^+\, \k_a,k^+_b\,\k_b,p_0^+,\p_0)&=&\frac{\rme^{i(k^-_a+k^-_b)(t_L-t_0)}  }{2p_0^+}\,\int_{\p_1,\q_1,\p_1'}\,(\lambda_a|i)\,(\lambda_b|j)\,(l|\lambda_0)\nn &\times&\ \int_{t_0}^{t_L} \rmd t_1\,(\k_a\,a;\k_b\,b|[{\cal G}(t_L,t_1)\cdot{\cal G}(t_L,t_1)]\,| \p_1\, a_1;\p_1'\,b_1)\nn
&\times& (\p_1\,a_1;\p_1'\,b_1|\, \Gamma^{ijl}\,|\q_1\,c_1)(\q_1\,c_1|\, {\cal G}(t_1,t_0)|\p_0\,c_0) ,\nn
\eeq
with $k^-_a=\k_a^2/(2k^+_a)$, $k^-_b=\k_a^2/(2k^+_b)$, $k_a^++k_b^+=p_0^+$,  and $(i|\lambda_a)$ denotes a transverse component of a polarization vector (recall that polarization is conserved in the propagation through the background field). A summation over repeated discrete indices is implied. 
We have defined
 \beq
&(\k_a\,a;\k_b\,b|\,[{\cal G}(t_L,t_1)\cdot {\cal G}(t_L,t_1)]\,| \p_1 \,a_1;\p_1'\,b_1)
\equiv (\k_a|{\cal G}^{aa_1}(t_L,t_1)\,| \p_1 )( \k_b  |\,{\cal G}^{bb_1}(t_L,t_1)\,| \p_1')\,,\nn
& (\p_1\,a_1\,i;\p_1'\,b_1\,j|\, \Gamma\,|\q_1\,c_1\,l)
\equiv (2\pi)^2\delta(\p_1+\p'_1-\q_1)\,2g\,f^{a_1b_1c_1}\Gamma^{ijl}(\p_1-z\q_1,z)\,,
 \eeq
 and we denote indifferently the matrix elements of the single propagator by 
$(\k_a|{\cal G}^{aa_1}(t_L,t_1)\,| \p_1 )$ or by $(\k_a\,a|{\cal G}(t_L,t_1)\,| \p_1\,a_1 )$. When the emitted gluon is soft, i.e. when it carries a small fraction $(1-z)$  of the energy of the parent gluon, we can use the following approximate expression for the vertex, with $z=k_a^+/p^+_0\lesssim 1$
\beq\label{eikonalGamma1}
\Gamma^{ijl}(\p-\q)\approx 
\frac{1}{1-z} (\p-\q)^j\delta^{li},
\eeq
 where $\p-\q$ is minus  the momentum of the soft gluon.
  
\begin{figure}[h]
\begin{center}\label{realandvirtual}
\includegraphics[scale=0.5]{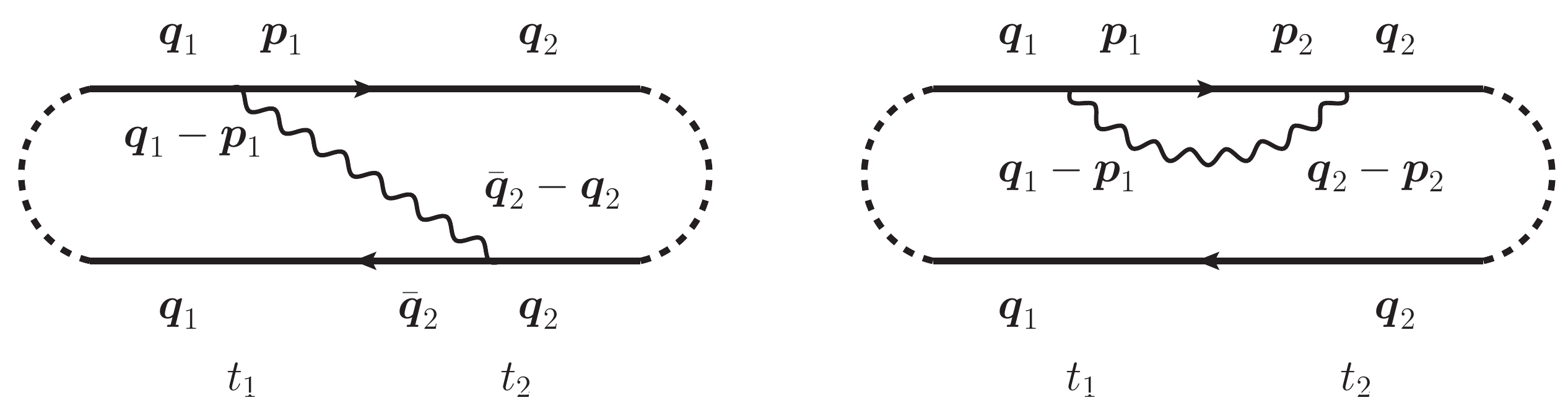}
\caption{The flow of momenta in the real and virtual contributions to the radiative correction to the 2-point function. The wavy line represents the soft radiated gluons, whose momenta flow from left to right. }
\end{center}
\end{figure}

 The calculation of the correction to the momentum broadening probability ${\cal P}$ due to one splitting consists in squaring the sum of the amplitude without splitting and that with one splitting, doing the appropriate sums and averages over colors and polarizations, integrating over the soft gluon. In doing the latter operation, the following identity is useful
 \beq 
\sum_b\int_{\k_b}(\p'_2|{\cal G}^{\dagger\,\bar b_2 b}(t_2,t_L)|\k_b)(\k_b|{\cal G}^{bb_2}(t_L,t_2)|\q'_2)=(2\pi)^2 \delta^{(2)}(\p'_2-\q_2')\delta_{\bar b_2 b_2}.
\eeq
Finally, we perform the average over the medium background field. The net result of this calculation can be expressed in the form $\Delta{\cal P}=\Delta{\cal P}_r+\Delta{\cal P}_v$, where $\Delta{\cal P}_r$ and $\Delta{\cal P}_v$ refer to the two diagrams displayed in Fig.~\ref{fig5} as real and virtual contributions. The figure \ref{realandvirtual} here makes explicit the flow of momenta in each contribution, and facilitates the reading of the formulae below. 

The  real contribution is given by 
\beq\label{Sigma2r}
&&\Delta {\cal P}_r(\k_a,\p_0;t_L,t_0)
=\frac{g^2N_c}{4\pi}\,2\text{Re} \,\int\frac{\rmd \omega_2}{\omega_2^3}\,\int_{t_0}^{t_L}\rmd t_2\int_{t_0}^{t_2}\rmd t_1\,\int_{\p_1\q_1\bar\q_2 \q_2} \; (\p_1-\q_1)\cdot (\bar\q_2-\q_2)\nn
&& \times\,{\cal P}(\k_a-\q_2;t_L-t_2)\tilde S^{(3)}(\q_1-\p_1,\bar\q_2-\q_2,\bar\q_2-\q_1;t_2,t_1){\cal P}(\q_1-\p_0;t_1-t_0), \nn
\eeq
where $\omega_2=(1-z)p^+_0$, and the factor $2\text{Re}$ accounts for the two time orderings of the times $t_1$ and $t_2$ at which the emission takes place in the amplitude and the complex conjugate amplitude, respectively. The function $\tilde S^{(3)}$ is the reduced 3-point function given in Eq.~(\ref{3-point-tilde}).

The virtual term is given by a similar expression
\beq\label{Sigma2r5}
&&\Delta{\cal P}_v=-\frac{g^2N_c}{4\pi}\,2\Re e\,\int
\frac{\rmd \omega_2}{\omega_2^3}\,\int_{t_0}^{t_L}\rmd t_2\int_{t_0}^{t_2}\rmd t_1\,\int_{ \p_1\q_1\p_2 \q_2} \,(\p_1-\q_1)\cdot(\q_2-\p_2)\,\nn
&& \times\,  {\cal P}(\k_a-\q_2;t_L-t_2) \tilde S^{(3)}(\q_1-\p_1,\q_2-\p_2, \q_2-\q_1;t_2,t_1) {\cal P}(\q_1-\p_0;t_1-t_0).\nn
\eeq
Here, the factor $2\text{Re}$ counts the two processes in which the virtual correction is attached to the amplitude (i.e. to a ${\cal G}$) or to its complex conjugate (i.e. to a ${\cal G}^\dagger$).

By adding the two contributions, one  obtains the modification to ${\cal P}$ due to one radiative correction. We get
\beq\label{deltaPapp}
&&\Delta{\cal P}(\k_a\, t_L|\p_0\, t_0)=\frac{g^2N_c}{4\pi}\,2\Re e\,\int
\frac{\rmd \omega_2}{\omega_2^3}\,\int_{t_0}^{t_L}\rmd t_2\int_{t_0}^{t_2}\rmd t_1\,\int_{ \q_1\q_2}{\cal P}(\k_a-\q_2;t_L-t_2)\nn
&&\qquad \times\,\left\{ \int_{ \p_1\bar\q_2} (\p_1-\q_1)\cdot (\bar\q_2-\q_2) \; \tilde S^{(3)}(\q_1-\p_1,\bar\q_2-\q_2,\bar\q_2-\q_1;t_2,t_1)-\right.\nn
&&\quad\qquad \left. \int_{\p_1\p_2}(\p_1-\q_1)\cdot(\q_2-\p_2)\; \tilde S^{(3)}(\q_1-\p_1,\q_2-\p_2, \q_2-\q_1;t_2,t_1) \right\}\nn
&&\quad\times\,{\cal P}(\q_1-\p_0;t_1-t_0)\,.
\eeq
Let us set $\l=\q_2-\q_1$. By changing  variables,  $\p_1-\q_1\rightarrow \q$, $\q_2-\p_2\rightarrow \q'$ in the second term, and $\bar\q_2-\q_2\rightarrow \q' $ in the first term (keeping $\q_1$ and $\q_2$ fixed), we can combine the terms within the curly brakets in the expression above, and obtains the expression used in Eq.~(\ref{deltaP}):
\beq\label{deltaP1}
\int_{ \q \q'}\, (\q\cdot\q')  \;\left[  \tilde S^{(3)}(\q,\q',\l+\q')-
\tilde S^{(3)}(\q,\q', \l)\right]\,. \nn
\eeq


\end{document}